\documentclass[aip,jcp,citeautoscript,preprint,noshowpacs,superscriptaddress,preprintnumbers,nofootinbib,showpacs,floatfix,12pt]{revtex4-2}

\usepackage{graphicx}
\usepackage{dcolumn}
\usepackage{bm}
\usepackage{comment}
\usepackage[utf8]{inputenc}
\usepackage[T1]{fontenc}
\usepackage{mathptmx}
\usepackage{placeins}
\usepackage{etoolbox}
\usepackage{hyperref}
\usepackage{epstopdf}
\usepackage{dsfont}

\usepackage{natbib}

\begin{document}

\title[]{Controlling Effective Hamiltonians: Broadband Pulsed Dynamic Nuclear Polarization by Constrained Random Walk and Non-linear Optimization}  

\author{Anders B. Nielsen}
\altaffiliation{Contributed equally} 
\affiliation{Interdisciplinary Nanoscience Center (iNANO) and Department of Chemistry, Aarhus University, Gustav Wieds Vej 14, DK-8000 Aarhus C, Denmark} 

\author{Jos{\' e} P. Carvalho}
\altaffiliation{Contributed equally} 
\affiliation{Interdisciplinary Nanoscience Center (iNANO) and Department of Chemistry, Aarhus University, Gustav Wieds Vej 14, DK-8000 Aarhus C, Denmark}

\author{Nino Wili}
\affiliation{Interdisciplinary Nanoscience Center (iNANO) and Department of Chemistry, Aarhus University, Gustav Wieds Vej 14, DK-8000 Aarhus C, Denmark} 

\author{Filip V. Jensen}
\affiliation{Interdisciplinary Nanoscience Center (iNANO) and Department of Chemistry, Aarhus University, Gustav Wieds Vej 14, DK-8000 Aarhus C, Denmark} 

\author{David L. Goodwin}
\affiliation{Interdisciplinary Nanoscience Center (iNANO) and Department of Chemistry, Aarhus University, Gustav Wieds Vej 14, DK-8000 Aarhus C, Denmark} 

\author{Thomas S. Untidt}
\affiliation{Interdisciplinary Nanoscience Center (iNANO) and Department of Chemistry, Aarhus University, Gustav Wieds Vej 14, DK-8000 Aarhus C, Denmark} 

\author{Zd{\v e}nek To{\v s}ner}
\affiliation{Department of Chemistry, Faculty of Science, Charles University in Prague, Hlavova 8, CZ-128 43, Czech Republic} 

\author{Niels Chr. Nielsen}
\email{ncn@chem.au.dk}
\affiliation{Interdisciplinary Nanoscience Center (iNANO) and Department of Chemistry, Aarhus University, Gustav Wieds Vej 14, DK-8000 Aarhus C, Denmark}




\begin{abstract}

We present constrained random walk (cRW) and figure of merit (FOM) based non-linear optimization procedures for systematic design and fundamental understanding of magnetic resonance experiments dressing bilinear and linear effective Hamiltonians to provide broadband polarization transfer. cRW can be used directly for fast random experiment design, or in combination with non-linear optimization or optimal control, leveraging the optimization of a FOM function for efficient control of linear and bilinear terms derived by exact effective Hamiltonian theory (EEHT). The efficacy of the combined cRW and FOM-based optimization approach is demonstrated by the design of broadband dynamic nuclear polarization (DNP) pulse sequences for static solids with an electron spin excitation bandwidth reaching 100 MHz.
 
\end{abstract}

\maketitle

\section{INTRODUCTION}

 The growing interest in quantum sensors probing nanoscale environments of unpaired electron spins, the availability of fast arbitrary waveform generators (AWGs) for pulsed microwave (MW) irradiation, and the proven ability of dynamic nuclear polarization (DNP) to boost the sensitivity nuclear magnetic resonance (NMR) drive strong interest in realizing pulsed DNP as alternative to current continuous wave (CW) DNP. From an NMR perspective, the vision is to add a pulsed electron spin MW channel to the already existing nuclear spin radio-frequency (RF) channels. From a reversed perspective, the increased use of nuclear spin RF channels may expand the scope of electron paramagnetic resonance (EPR) and hyperfine spectroscopy. This blending of pulsed operation on electron and nuclear spins challenges the design of pulse sequences, as the magnitude of the interactions involving electron spins are orders of magnitude larger than NMR-relevant nuclear spin interactions, and the time window available for operation is orders of magnitude shorter than encountered in typical NMR pulse sequences. In addition, the size of the involved spin system may grow enormously due to the nm distance range of electron-spin involved interactions, as opposed to \AA \ length scales for NMR-relevant nuclear spin interactions. 

The advancement of liquid- and solid-state NMR,\cite{Ernst_book} magnetic resonance imaging (MRI), and EPR\cite{Jeschke_book} have relied on parallel development of advanced instrumentation and sophisticated application-oriented methods designed using powerful theoretical and numerical methods. The design toolbox includes analytical tools such as average or effective Hamiltonian,\cite{AHT,MaricqWaugh,scBCH}  product operator,\cite{Prodop} Floquet,\cite{Shirley,Weintraub,Scholz2010} exact effective Hamiltonian (EEHT),\cite{EEHT,EEHT2} and single-spin vector effective Hamiltonian (SSV-EHT)\cite{shankar,SSV-EHT,SVEHT_EEHT} theories, often in combination with numerical calculations,\cite{SIMPSON,spinevolution,spinach} non-linear optimization,\cite{press2007numerical} and optimal control.\cite{Pontryagin,Krotov,Morzhin_2019,OC1,grape,OCsolid,MaximovOC,OCNCONCA,SIMPSONOC,MaximovSmoothing,OCMRI_Krotov,OCKuprov,Tosner:2018tb,TosnerSciAdv} These tools have  enabled design of thousands of advanced pulse sequences extracting detailed atomic resolution structure and dynamics information from increasingly complex molecular systems in areas spanning from materials science to biology and medicine. The same tools, taking intuitive inspiration from solid-state NMR and EPR, are gradually finding increasing use also for the development of pulsed DNP experiments such as NOVEL,\cite{NOVEL} TOP,\cite{TOP_DNP} BEAM,\cite{BEAM} XiX,\cite{XiX_DNP} TPPM,\cite{TPPM_DNP} and PLATO.\cite{PLATO_adv} The use of non-linear optimization and optimal control has also been shown to be a promising avenue for design of DNP experiments tailored to tackle the complexity electron-nuclear spin systems in different contexts.\cite{MaximovOC,OC_MRI_DNP,PLATO_adv,OC_DNP} 

In this work, we take further steps along the line of effective Hamiltonian-based approaches to non-linear optimization and optimal control, which may lead to better control and understanding of the spin dynamics in large spin systems than prevailing state-to-state optimization protocols.\cite{OC_DNP} This includes the introduction of a simple constrained random walk (cRW) approach to design multiple-pulse dipolar recoupling experiments by matching resonances between external effective fields as in DNP and/or macroscopic sample rotation as in solid-state NMR. This may provide an important new tool in the spin engineering toolbox either by itself for fast random development of dipolar recoupling experiments without optimization. It  may also be used to remedy the long-standing problem of finding good initial guesses for optimal control design procedures. The latter become particularly important when only a few resonance conditions exist, as is often the case for DNP as opposed to manifolds of resonances as typically encountered in magic-angle-spinning (MAS) dipolar recoupling. 

We address, in detail, the need to appropriately balance linear and bilinear fields to provide efficient polarization transfer through dipolar recoupling covering experimental conditions with large variation in linear external fields like resonance offsets or variations in radio-frequency (RF) or microwave (MW) amplitudes due to field inhomogeneity. We note that we here extend the notation of dipolar recoupling as well-known in solid-state NMR to recouple dipole-dipole couplings under MAS conditions to also cover "recoupling" of pseudo-secular hyperfine couplings in presence of a large Zeeman interaction on nuclear spins as encountered in DNP. Efficient dipolar recoupling is achieved through optimization of a figure of merit (FOM) function to balance relevant terms in the effective Hamiltonian, here formulated via Exact Effective Hamiltonian Theory (EEHT). As a fundamental element, it is demonstrated that deliberate design of experiments with sizable linear fields may be used actively to truncate the effective Hamiltonian to be operative in invariant zero- or double-quantum 3D operator subspaces ideally served for polarization transfer by planar ZQ or DQ operators. This is an important new component in effective Hamiltonian-based pulse sequence design so far mainly relying on concatenation of pulse sequences elements with different phases. Here we let the element truncate its own effective Hamiltonian just by repetition. Lastly, we combine the cRW and FOM-based non-linear optimization into a design protocol for dipolar recoupling and demonstrate its features theoretically, numerically, and experimentally in context of broadband DNP experiments reaching 100 MHz offset compensation on the electron spins, using a peak microwave (MW) amplitude of 32 MHz. 

\section{THEORY}

We will in this section address theoretically the two aspects introduced above, namely (1) the constrained Random Walk (cRW) method as a stand-alone tool to develop recoupling experiments or as a supplementary tool to establish educated starting points for non-linear optimization or optimal control and (2) the use of FOM-based optimization to actively balance linear and bilinear effective fields in optimization of dipolar recoupling experiments with focus on pulsed DNP on static samples. These two aspects, including the important aspect of internal truncation of effective Hamiltonians by sizable linear effective fields, are incorporated into a framework of numerical optimization of DNP pulse sequences providing efficient broadband transfer of polarization from unpaired electron spins to nuclear spins in static samples. We note that these aspects may have direct analog to dipolar recoupling in MAS solid-state NMR which will briefly be addressed at relevant places in the text without adding the complexity of carrying additional radio-frequency (RF) and time-modulation of anisotropic interactions into the theory presented here.

\subsection{The Hamiltonian and propagation}

With the focus on design of DNP experiments, we consider the Hamiltonian for an electron (S) - nuclear (I) spin-pair Hamiltonian in the rotating frame of the S spins 
\begin{equation}
  \mathcal{H}(t)=\omega_{0I}  I_z+\Delta\omega_S  S_z+A S_z I_z+B S_z I_x+\mathcal{H}_{\text{MW}}(t)
  \label{Eq:1}
\end{equation}  
with frequencies given in angular units. The first two terms describe the $I$-spin Zeeman operator, with $\omega_{0I}$ being the nuclear spin Larmor frequency, and the electron spin offset  $\Delta\omega_{\text{S}}=\omega_{\text{S}}-\omega_{\text{MW}}^{\text{carrier}}$ relative to the MW carrier frequency $\omega_{\text{MW}}^{\text{carrier}}$. The next two terms describe the secular and pseudo-secular hyperfine coupling with amplitudes $A$ and $B$, respectively. The last  term accounts for time-dependent MW  irradiation on the electron spins. We note that RF irradiation on the nuclear spins is ignored due to the large time- and amplitude scale differences between electron and nuclear spin dynamics. The pulsed MW irradiation, formulated as
\begin{equation}
  \mathcal{H}_{\text{MW}}(t)=\omega_{S,x}^{\text{MW}}(t)  S_x + \omega_{S,y}^{\text{MW}} (t) S_y  ,
  \label{Eq:2}
\end{equation}
with $x-$ and $y$-phase amplitudes (control fields) $\omega_{S,x}^{\text{MW}}(t)$ and $\omega_{S,y}^{\text{MW}}$, respectively, may drive polarization transfer through the pseudo-secular coupling term. In the analogous solid-state NMR case, RF irradiation is needed on both spins to drive polarization transfer through the secular term, typically reformulated with the operator being $2S_zI_z$ instead of $S_zI_z$ and $A(t)/2=\omega_{IS}(t)$ where we introduced time dependency for the coupling ($\omega_{IS}(t)$) invoked by sample rotation or transformation into the interaction frame of the pulse sequence. 

The transfer of polarization is characterized by the density operator evolution
\begin{equation}
  \rho(t)=U(t) \rho(0) U^\dagger(t)  ,
  \label{Eq:3}
\end{equation}
where the initial operator, $\rho(0)$, may be $S_z$ or $S_x$ (initialized by a $y$-phase $\pi/2$ pulse), and the propagator given by
\begin{equation}
  U(t)=U_n U_{n-1} .... U_2 U_1,
  \label{Eq:4}
\end{equation}
considering a pulse sequence with $n$ pulses each of which characterized by
\begin{equation}
  U_j= \hat{T} \exp[-i\int_{t_{j-1}}^{t_{j}}H_j(t)dt] ,
  \label{Eq:5}
\end{equation}
relating directly to the Hamiltonian in Eq. (\ref{Eq:1}). A typical aim in pulse sequence design is to have  the control fields in $\mathcal{H}_{\text{MW}}(t)$, within the amplitude limits of available instrumentation, adjusted to reach  maximal polarization transfer within a given mixing time $t_M$. In a typical GRAPE-\cite{grape} or Krotov-\cite{MaximovOC} based optimal control approach, the objective may be to optimize the state-to-state fidelity
\begin{equation}
  \mathcal{F}_{ss}(t_M)= 
  \frac{\langle  \rho_D(t_M) |  \rho(t_M)\rangle}{\langle  \rho_D(t_M) |  \rho_D(t_M)\rangle} ,
  \label{Eq:6}
\end{equation}
 where the density operator $\rho_D(t_M)$ is proportional to the spin operator to which transfer is desired. A potential challenge, in context of DNP experiment design, is appropriate formulation of the spin system to avoid exploitation of interactions which are not representative for the spin dynamics in large electron-nuclear spin systems as typically encountered in DNP. 
 
 To reduce the problem of finding representative (large) spin systems, it appears rational to optimize instead the effective Hamiltonian responsible for polarization transfer. This may be done indirectly by optimizing the match between the propagator delivered by the pulse sequence relative to a desired propagator $U_D(t_M)$ representing the desired effective Hamiltonian, \cite{OCEffHam} i.e.,
\begin{equation}
  \mathcal{F}_{U}(t_M)=\langle  U_D(t_M) |  U(t_M) \rangle ,
  \label{Eq:7}
\end{equation}
with the benefit of focusing on effective Hamiltonians which by their one- to two-spin nature may cope better with large spin systems. A challenge with this effective-propagator approach is that it is difficult to penalize undesired Hamiltonian components through the exponential operator representation in Eq. (\ref{Eq:7}) as discussed in Ref. \cite{OCEffHam} in the context of planar and isotropic mixing for solid-state MAS NMR polarization transfer. Additionally, it may involve certain \textit{à priori} constraints on the coefficients of the terms of the effective Hamiltonian, that may limit its general adaptability as well as prevent it from exploring the full parameter space.

\sloppy{
With direct optimization of the effective Hamiltonian not being straightforward, we have within the framework of the SSV-EHT formalism \cite{shankar,SSV-EHT,SVEHT_EEHT,BEAM} proposed the optimization of periodic sequences of length $t_m$ repeated a number to times $n$ to give a total transfer time of $t_M=n t_m$.\cite{OC_DNP} The simplest approach is a replicated state-to-state OC method, where the repetition of pulse sequence building block of lengt $t_m$ is imposed via the matrix-power method. Alternatively,  a logarithmic mapping \cite{PLATO_adv,OC_DNP} can be used to optimize a so-called figure-of-merit (FOM) function }
\begin{equation}
  \mathcal{F}_{\text{FOM}}(t_M)=\frac{\langle  \mathcal{O}_D |  \rho_{\text{FOM}}(t_M)\rangle}{\langle  \mathcal{O}_D |  \mathcal{O}_D\rangle} .
  \label{Eq:8}
\end{equation}
This is essentially also a density operator projection onto a desired operator $ \mathcal{O}_D$ within an optimal time $t_M$, but here the density operator $\rho_{\text{FOM}}(t_M)$ is calculated with the effective Hamiltonian for the building block (of duration $t_m$)  `filtered' to contain only elements belonging to zero- and/or double-quantum (ZQ or DQ) subspaces. Note that this `filtration', as well as the replication of a pulse sequence building block via the matrix power, may only be useful under conditions of pulse sequences with effective fields stronger than the hyperfine coupling as elaborated further on below. To unambiguously define the involved ZQ and/or DQ operator terms these approaches may benefit from transformation into an operator frame oriented with the effective field of the MW pulse sequence along the longitudinal axis, as invoked by an SU(2) unitary transformation
\begin{equation}
  \tilde{S}_i = V^\dagger S_i V, 
  \label{Eq:9}
\end{equation}
with $i=x,y,z$ and $\tilde{S}$ representing the tilted frame, and $V$ the unitary transformation operator.  Formulas for and derivation of this transformation may be found in Ref. (\cite{OC_DNP}).

\subsection{EEHT formulation of FOM optimization}

In the case of MW pulse sequences with an effective field exceeding the hyperfine coupling, it is meaningful to establish a FOM fidelity function to optimize 
 $\tilde{S}_z$ to $I_z$ transfer in the effective frame of the electron spin MW irradiation. To first order this function may be formulated as
\begin{equation}
\mathcal{F}_{\text{FOM}}^p(t_M)=
\pm \langle  \rho(0) | {\tilde S}_z\rangle
  \frac{|\omega_{\text{bil}}^p|^2}{|\omega_{\text{bil}}^p|^2+4|\omega_{\text{lin}}^p|^2}
  \sin^2\left[\frac{t_M}{4}\sqrt{|\omega_{\text{bil}}^p|^2+4|\omega_{\text{lin}}^p|^2} \right] ,
  \label{Eq:10}
\end{equation}
where $t_M = n t_m$ is the overall mixing time during which MW irradiation is invoked by repeating a basic pulse sequence element of duration $t_m$ $n$ times to achieve maximum transfer. The prefactor $\langle  \rho(0) | {\tilde S}_z\rangle$ imposes appropriate alignment of the initial density operator relative to the direction of the effective field of the MW pulse sequence (which in the tilted frame is directed along $z$). We assume in this study $\rho(0)=S_x$. 

For ZQ planar mixing ($p$=ZQ, positive transfer marked by + sign) the frequency components, calculated for the pulse sequence element (duration $t_m$), is given by 
\begin{eqnarray}
|\omega_{\text{lin}}^{ZQ}| &=& |\omega_{\tilde{S}z}-\omega_{Iz}|
  \label{Eq:11}
\\
|\omega_{\text{bil}}^{ZQ}| &=& \sqrt{(\omega_{\tilde{S}xIx}+\omega_{\tilde{S}yIy})^2
+
(\omega_{\tilde{S}yIx}-\omega_{\tilde{S}xIy})^2 ,
  \label{Eq:12}
} 
\end{eqnarray}
while the corresponding terms for DQ planar mixing ($p$=DQ; negative transfer marked by $-$ sign) take the form
\begin{eqnarray}
|\omega_{\text{lin}}^{DQ}| &=& |\omega_{\tilde{S}z}+\omega_{Iz}|
  \label{Eq:13}
\\
|\omega_{\text{bil}}^{DQ}| &=& \sqrt{(\omega_{\tilde{S}xIx}-\omega_{\tilde{S}yIy})^2
+
(\omega_{\tilde{S}yIx}+\omega_{\tilde{S}xIy})^2
} .
  \label{Eq:14}
\end{eqnarray}
The single-spin (linear, 'lin') and two-spin (bilinear, 'bil') are given by projections of the effective Hamiltonian in the effective-field frame ($\overline{\tilde{\mathcal{H}}}$) onto the relevant operators as
\begin{eqnarray}
\omega_{Qz}&=&\langle  Q_z | \overline{\tilde{\mathcal{H}}}\rangle 
  \label{Eq:15}
\\
\omega_{\tilde{S}_pI_q}&=&\langle  \tilde{S}_pI_q | \overline{\tilde{\mathcal{H}}}\rangle
  \label{Eq:16}
\end{eqnarray}
where $Q$ = $I$ or $\tilde{S}$.

The effective Hamiltonian may be analytically derived using Exact Effective Hamiltonian Theory (EEHT)\cite{EEHT,EEHT2}
\begin{equation}
\overline{\tilde{\mathcal{H}}}=\frac{1}{-it_m}\ln{\tilde{U}(t_m)} = \frac{1}{-it_m} \sum_{i=0}^{N-1}g_i \tilde{X}(t_m)^i ,
\label{Eq:17}
\end{equation}
where $N$ is the degree of the special unitary group SU($N$) and 
\begin{equation}
\tilde{X}(t_m)=E-\tilde{U}(t_m) ,
\label{Eq:18}
\end{equation}
with $E$ being the unity operator and $\tilde{U}(t_m)$ the propagator (cf. Eq. (\ref{Eq:4})) for the pulse sequence element in the tilted frame. The $g_i$ coefficients may be determined from the eigenvalues of $\tilde{X}(t_m)$. For SU(2) this gives
\begin{eqnarray}
g_0&=&\frac{1}{m}[\lambda_1 \ln (1-\lambda_2)-\lambda_2 \ln (1-\lambda_1)] 
\label{Eq:19} \\
g_1&=&\frac{1}{m}[\ln (1-\lambda_1)-\ln (1-\lambda_2)] ,
\label{Eq:20}
\end{eqnarray}
with $m=\lambda_1-\lambda_2$. The SU(2) representation may conveniently be used for establishment of the frame transformation in Eq. (\ref{Eq:9}). For SU(4), in the typical case of non-degenerated eigenvalues, the corresponding coefficients become
\begin{eqnarray}
g_0 &=& \frac{1}{m}[\ln (1-\lambda_4) \left(\lambda_1^3
\left(\lambda_2^2 \lambda_3-\lambda_2 \lambda_3^2\right)+\lambda_2^3
\left(\lambda_1 \lambda_3^2-\lambda_1^2 \lambda_3\right)+\lambda_3^3
\left(\lambda_1^2 \lambda_2-\lambda_1 \lambda_2^2\right)\right) \nonumber \\
&+&\ln(1-\lambda_3) \left(\lambda_1^3 \left(\lambda_2 \lambda_4^2-\lambda_2^2
\lambda_4\right)+\lambda_2^3 \left(\lambda_1^2 \lambda_4-\lambda_1
\lambda_4^2\right)+\lambda_4^3 \left(\lambda_1 \lambda_2^2-\lambda_1^2
\lambda_2\right)\right) \nonumber \\
&+&\ln (1-\lambda_2) \left(\lambda_1^3
\left(\lambda_3^2 \lambda_4-\lambda_3 \lambda_4^2\right)+\lambda_3^3
\left(\lambda_1 \lambda_4^2-\lambda_1^2 \lambda_4\right)+\lambda_4^3
\left(\lambda_1^2 \lambda_3-\lambda_1 \lambda_3^2\right)\right) \nonumber \\
&+&\ln
(1-\lambda_1) \left(\lambda_2^3 \left(\lambda_3 \lambda_4^2-\lambda_3^2
\lambda_4\right)+\lambda_3^3 \left(\lambda_2^2 \lambda_4-\lambda_2
\lambda_4^2\right)+\lambda_4^3 \left(\lambda_2 \lambda_3^2-\lambda_2^2
\lambda_3\right)\right)] \label{Eq:21}
\end{eqnarray}
\begin{eqnarray}
g_1&=&\frac{1}{m}[\ln (1-\lambda_4) \left(\lambda_1^3
\left(\lambda_3^2-\lambda_2^2\right)+\lambda_2^3
\left(\lambda_1^2-\lambda_3^2\right)+\lambda_3^3
\left(\lambda_2^2-\lambda_1^2\right)\right) \nonumber \\ 
&+&\ln (1-\lambda_3)
\left(\lambda_1^3 \left(\lambda_2^2-\lambda_4^2\right)+\lambda_2^3
\left(\lambda_4^2-\lambda_1^2\right)+\lambda_4^3
\left(\lambda_1^2-\lambda_2^2\right)\right) \nonumber \\
&+&\ln (1-\lambda_2)
\left(\lambda_1^3 \left(\lambda_4^2-\lambda_3^2\right)+\lambda_3^3
\left(\lambda_1^2-\lambda_4^2\right)+\lambda_4^3
\left(\lambda_3^2-\lambda_1^2\right)\right) \nonumber \\
&+&\ln (1-\lambda_1)
\left(\lambda_2^3 \left(\lambda_3^2-\lambda_4^2\right)+\lambda_3^3
\left(\lambda_4^2-\lambda_2^2\right)+\lambda_4^3
\left(\lambda_2^2-\lambda_3^2\right)\right)]
\label{Eq:22}
\end{eqnarray}
\begin{eqnarray}
g_2&=&\frac{1}{m}[\ln (1-\lambda_4) \left(\lambda_1^3
(\lambda_2-\lambda_3)+\lambda_2^3 (\lambda_3-\lambda_1)+\lambda_3^3
(\lambda_1-\lambda_2)\right) \nonumber \\
&+&\ln (1-\lambda_3) \left(\lambda_1^3
(\lambda_4-\lambda_2)+\lambda_2^3 (\lambda_1-\lambda_4)+\lambda_4^3
(\lambda_2-\lambda_1)\right) \nonumber \\
&+&\ln (1-\lambda_2) \left(\lambda_1^3
(\lambda_3-\lambda_4)+\lambda_3^3 (\lambda_4-\lambda_1)+\lambda_4^3
(\lambda_1-\lambda_3)\right) \nonumber \\
&+&\ln (1-\lambda_1) \left(\lambda_2^3
(\lambda_4-\lambda_3)+\lambda_3^3 (\lambda_2-\lambda_4)+\lambda_4^3
(\lambda_3-\lambda_2)\right)] \label{Eq:23}
\end{eqnarray}
\begin{eqnarray}
g_3&=&\frac{1}{m}[\ln (1-\lambda_4) \left(\lambda_1^2
(\lambda_3-\lambda_2)+\lambda_2^2 (\lambda_1-\lambda_3)+\lambda_3^2
(\lambda_2-\lambda_1)\right)\nonumber \\
&+&\ln (1-\lambda_3) \left(\lambda_1^2
(\lambda_2-\lambda_4)+\lambda_2^2 (\lambda_4-\lambda_1)+\lambda_4^2
(\lambda_1-\lambda_2)\right)\nonumber \\
&+&\ln (1-\lambda_2) \left(\lambda_1^2
(\lambda_4-\lambda_3)+\lambda_3^2 (\lambda_1-\lambda_4)+\lambda_4^2
(\lambda_3-\lambda_1)\right)\nonumber \\
&+&\ln (1-\lambda_1) \left(\lambda_2^2
(\lambda_3-\lambda_4)+\lambda_3^2 (\lambda_4-\lambda_2)+\lambda_4^2
(\lambda_2-\lambda_3)\right)] \label{Eq:24}
\end{eqnarray}
using
$m=(\lambda_1-\lambda_2) (\lambda_1-\lambda_3) (\lambda_1-\lambda_4)
(\lambda_2-\lambda_3) (\lambda_2-\lambda_4) (\lambda_3-\lambda_4)$. We note that all cases in this study were covered by this non-degenerated eigenvalue case. Otherwise, formulas for the degenerated cases can be found in Ref. (\cite{EEHT}). We note that the pertinent frequencies in Eq. (\ref{Eq:10}) are calculated for the sequence element of length $t_m$ using the SU(4) EEHT formulation, while the full excitation is optimized for multiple repetitions of the pulse sequence element $t_M = n t_m$. The optimal value of the integer $n$ depends on the size of $|\omega_{\text{bil}}^p|$ and thereby on the pseudo-secular hyperfine coupling $B$ and the scaling factor of the bilinear terms imposed by the MW pulse sequence element.

\subsection{Truncation of the effective Hamiltonian by effective fields}

It is important to realize that the EEHT (or alternative matrix log) formulation of the effective Hamiltonian is completely general and provides the projection of all 16 terms (including the identity) of the two-spin-1/2 effective Hamiltonian. Among these are 6 linear terms (3 Cartesian components for each spin) which in the effective field frame have finite coefficients only for the longitudinal terms only (i.e., $\tilde{S}_z$ and $I_z$). Provided these effective linear fields by design are chosen to be sufficiently large, they truncate a good portion of the 9 bilinear terms, which all scales with the hyperfine coupling $B$ through the pseudo-secular coupling term in Eq. (\ref{Eq:1}) and a scaling factor ($\leq$ 1) induced by the pulse sequence element. 

In proximity of resonance, the $I$- and $S$-spin effective fields are similar and the condition for efficient truncation may be formulated as $\omega_{\text{eff}}^{(S)} \approx \omega_{\text{eff}}^{(I)} > B$ (\textit{vide supra}). In this case, through rotation of the effective Hamiltonian by the effective longitudinal fields $\omega_{\text{eff}}^{(I)} I_z$ and $\omega_{\text{eff}}^{(S)} \tilde{S}_z$, all bilinear terms containing only one transverse operator component will be truncated to first order upon repetitive application of the pulse sequence element (recommended $n$ being 3 or more). We note that this truncation, beyond $n$, depends on the scaling of the pseudo-secular hyperfine coupling by the pulse sequence. This implies that lower effective fields may be required for truncation for MW pulse sequences with low scaling for the bilinear terms. The $I$-spin effective field truncates the effective Hamiltonian components $I_q\tilde{S}_z$ and the $S$-spin effective field truncates the effective Hamiltonian components $I_z\tilde{S}_q$ with $q=x$ or $y$, effectively leaving the 4 operator terms with $I_q\tilde{S}_r$ with $q,r=x$ or $y$ accessible for dipolar recoupling upon  matching of the effective fields. The $I_z\tilde{S}_z$ term cannot contribute to recoupling, as the transfer of polarization is mediated through $\tilde{S}_z$ to $I_z$ transfer in the frame of the effective fields (RF irradiation is required to activate this term). The remaining active effective Hamiltonian components may be arranged in ZQ or DQ invariant fictitious spin-1/2\cite{fictitious_vega,fictitious_wokaun_ernst} 3D operator spaces, depending on the choice of resonance. We note that the truncation of the effective Hamiltonian does not depend on establishment of resonance, it relies only on the size of the individual I or S spin effective fields relative to the effective pseudo-secular hyperfine coupling. Upon satisfying truncation, the spin dynamics in dipolar recoupling may be described, to a good approximation, in terms of the FOM-based fidelity function given in Eq. (\ref{Eq:10}). We will in the Results and Discussion section provide an illustration of this fundamentally important truncation (henceforth referred to as auto-truncation) aspects discussed above. 

\subsection{Resonance condition}

\begin{figure}
    \centering
    \includegraphics[width=0.5\textwidth]{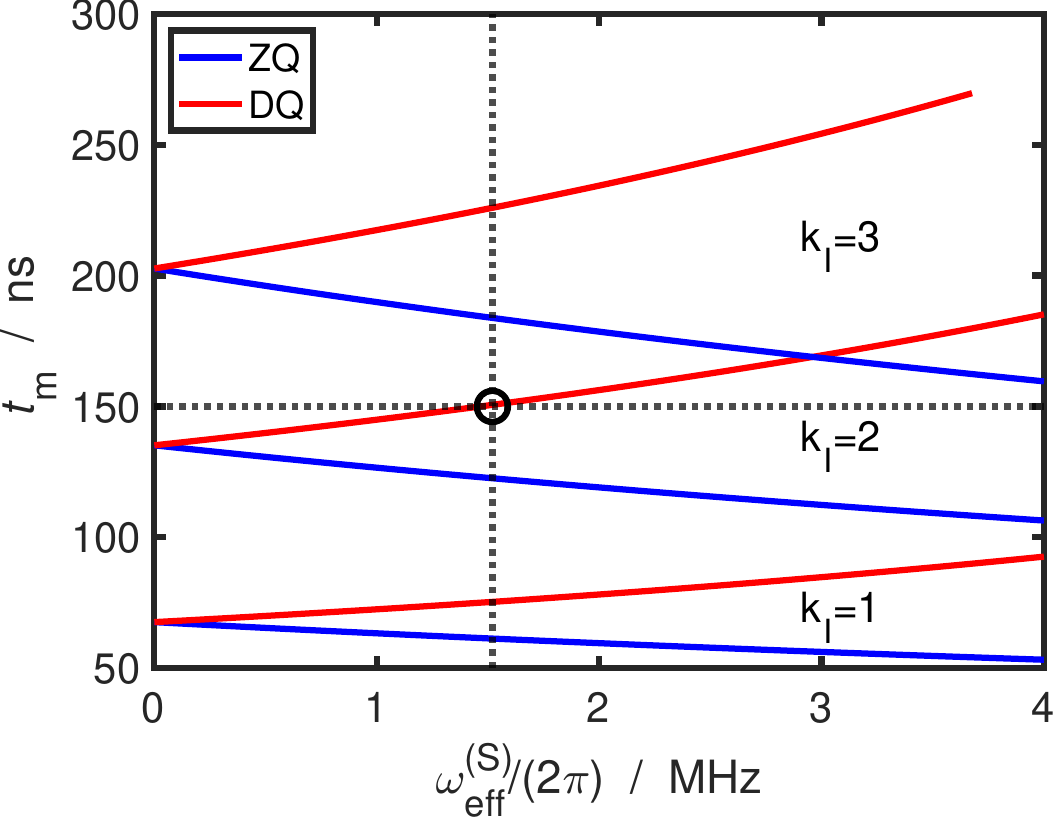}
    \caption{Dipolar recoupling ZQ (blue) and DQ (red) resonance conditions for DNP polarization transfer as function of the MW pulse sequence effective field (horizonthal) and the modulation time (duration of pulse sequence element, vertical) under condition of $\omega_{0I}/(2\pi)$ = 14.8 MHz. Note the missing red line in the top upper corner, reflecting $\omega_{\text{eff}}^S t_m \geq 2\pi$ which goes into the $k_I$=1 resonance upon subtracting $2 \pi$ from the rotation. The dotted line and the circle in black designate the resonance condition selected in this work for DNP pulse sequence optimization (see text).}
    \label{fig1}
\end{figure}

Equipped with analytical tools to evaluate the effective Hamiltonian for any MW pulse sequence in terms of effective fields and upon truncation by effective fields relevant contributions to ZQ- and/or DQ-subspace linear and bilinear Hamiltonians, it is possible to systematically design (and analyze) DNP pulse sequences. For DNP polarization transfer to occur it is necessary to fulfill the resonance condition
\begin{equation}
\omega_{0I} - k_I \omega_m \pm \omega_{\text{eff}}^{(S)} = 0 , \label{Eq:25}
\end{equation}
bearing dependencies on the Larmor frequency  $\omega_{0I}$ of the nuclear spins to which transfer is aimed at, an integer $k_I$,  the pulse sequence modulation frequency $\omega_m$, and the $S$-spin effective field $\omega_{\text{eff}}^{(S)}$ induced by the MW pulse sequence in the electron-spin effective-field frame. The sign in front of the $S$-spin effective field corresponds to ZQ ($-$) or DQ (+) transfer. The modulation frequency relates to the duration of the pulse sequence element as $\omega_m=2\pi/t_m$. To appreciate the details of the components in Eq. (\ref{Eq:25}), Fig. \ref{fig1} depicts for the case of X-band MW irradiation with $\omega_{0I}/(2\pi)$ = 14.8 MHz for $^1$H the ZQ (blue) and DQ (red) resonances as function of $k_I$, $\omega_{\text{eff}}^{(S)}$, and $t_m$.

Several important features should noticed: ($i$) The position and proximity of ZQ and DQ resonances depend on the modulation time and the size of the $S$-spin effective field. This is important, as too close resonances may provide destructive overlap of ZQ and DQ resonances transferring polarization with opposite sign, inducing risk of destructive interference. This risk may increase in case of MW inhomogeneity. ($ii$) The effective field relates to the amplitude of the applied MW field, implying that the available region of  resonances may be limited by the capability of  available MW amplifiers. Furthermore, the susceptibility to MW field inhomogeneity increases with the overall rotation angle induced by the effective field, implying that sequences with low effective field, yet avoiding overlap of ZQ and DQ resonances, may be most attractive. However, the effective field has to be sufficiently strong to ensure truncation of bilinear effective Hamiltonians with only one transverse component to first order and thereby maintain the relevant spin dynamics within invariant ZQ or DQ subspaces (i.e., $\omega_{\text{eff}}^{(S)} > B$).  ($iii$) The easily adjustable modulation time of the pulse sequence $t_m$  (i.e., the length of pulse sequence element) plays an important role in controlling the establishment of a good resonance condition. ($iv$) The product of the effective field and the modulation time represents the overall nutation angle of the pulse sequence which has a periodicity of 2$\pi$ implying that pulse sequences may jump between resonances as discussed in relation to Fig. \ref{fig1}. We note that the resonance condition in Eq. (\ref{Eq:25}) is a generalized case of the Hartmann-Hahn condition\cite{HartmanHahn} widely used for solid-state NMR cross-polarization and MAS dipolar recoupling. The resonance condition can, as demonstrated recently,\cite{DNP_SSNMR_Carvalho:2025aa}  easily be extended to cover MAS heteronuclear dipolar recoupling by matching the effective field optimized for the $S$-spin channel with an RF field $\omega_{I}^{\text{RF}}$ on the $I$-spin channel as
\begin{equation}
\omega_{I}^{\text{RF}} - k_I \omega_m \pm \omega_{\text{eff}}^{(S)} = n \omega_r , \label{Eq:26}
\end{equation}
 with $k_I$ and $n$ being integers, $\omega_r$ the angular MAS frequency, and  $\omega_m$ and $\omega_r$ chosen to ensure periodicity. This relationship implies that it is simple to translate pulse sequences design for one purpose (e.g., static sample DNP) to another (e.g., MAS dipolar recoupling).\cite{DNP_SSNMR_Carvalho:2025aa} We note that the constant field on the $I$-spin RF channel may obviously be replaced by a time-modulated RF pulse sequence instead. In this case,  the effective field $\omega_{\text{eff}}^{(I)}$ of the $I$-spin channel can be used in the resonance equation instead of $\omega_I^{\text{RF}}$.

\subsection{Constrained Random Walk (cRW)}

With the FOM function and the resonance conditions established, it is possible to systematically develop DNP pulse sequences offering efficient transfer of polarization with robustness towards resonance offsets and MW (and/or RF) inhomogeneity. This may be accomplished using optimal control procedures,\cite{OC_DNP} or in the case of repetitive pulse sequence elements with a modest number of pulse sequence variables using conventional non-linear optimization.\cite{PLATO_adv} A general challenge in such optimizations is the complexity of the fidelity function multi-parameter landscape with the potential of getting trapped in local extrema rather than reaching the global extreme. Despite a larger number of possible resonances (c.f., Eqs. (\ref{Eq:25}) and (\ref{Eq:26})), the same situation is commonplace in optimal control design of dipolar recoupling experiments in solid-state MAS NMR. Although  the determinants for dipolar recoupling as formulated above are quite simple, no general solution has to our knowledge been presented to efficiently cope with the challenge of establishing good starting guesses for numerical optimization of dipolar recoupling experiments. Neither has it been proposed to generate multiple-pulse dipolar recoupling experiments without involvement of advanced theoretical analysis or numerical optimization. Here we present  a  solution to both aspects, henceforth referred to as constrained Random Walk (cRW). The basic idea is quite simple. In any dipolar recoupling experiment, the aim is to ensure that the effective MW or RF fields integrated over time fulfill the matching conditions in Eqs. (\ref{Eq:25}) and (\ref{Eq:26}) in the context of DNP and heteronuclear MAS NMR polarization transfer, respectively.

While the formalism above should enable a general formulation in terms of pulse sequences with arbitrary phase and amplitudes, we consider here the most simple solution involving only $\pm x$-phase MW (or RF) irradiation. In this case, the accumulated evolution angles (see feature ($iv$) above) which has to be matched for  dipolar recoupling to occur are subject to the relations
\begin{eqnarray}
 \theta^{DNP} &=& \int_{0}^{t_m} \omega_{S}^{\text{MW}}(t) dt = \pm (\omega_{0I} t_m -k_I 2\pi) \quad ,
  \label{Eq:27} \\
\theta^{NMR} &=& \int_{0}^{t_m} \omega_{I}^{\text{RF}}(t) dt = n \omega_r t_m +k_I 2\pi \mp \int_{0}^{t_m} \omega_{S}^{\text{RF}}(t)dt \quad ,
  \label{Eq:28}
  \end{eqnarray}
for ZQ (upper sign) and DQ (lower sign) transfer in DNP and MAS NMR with $|n|$ = 1,2, respectively. $k_I$ is an integer ensuring the solution to be represented modulo 2$\pi$. This represent a generalization of  well-known cases with constant amplitude MW or RF irradiation. For DNP, the $\omega_{S}^{\text{MW}} = \omega_{I}$ match is known as the NOVEL condition.\cite{NOVEL} For solid-state NMR, the $\omega_{I}^{\text{RF}} = \omega_{S}^{\text{RF}}+n \omega_r$ match is known as the MAS adjusted Hartmann-Hahn  condition\cite{HartmanHahn} being the foundation of cross-polarization (CP) MAS NMR experiments. Equations (\ref{Eq:27}) and (\ref{Eq:28}) formulate the simple requirement for all (here exemplified by $x$-phase) recoupling experiments that the only things that matters are ($i$) that the accumulated rotation angle $\theta$ at the end of the pulse sequence element (modulo 2$\pi$) is the same on the two involved Larmor frequency/MW/RF channels potentially corrected for an integral number of accumulated MAS rotation angles and ($ii$) that the maximum MW (or RF) amplitudes are within the limits provided by available instrumentation. Within these limitations, the path in-between the starting point $t$=0 and the ending point $t=t_m$ does not matter in regard of achieving recoupling. 

With the initial (trivial) and ending points (overall length of the pulse sequence element, $t_m$ and the overall accumulated angle $\theta(t_m)$) constrained, the cRW procedure may be implemented to select the pulse duration and amplitudes sequentially, gradually imposing more and more constraints on the subsequent pulses to reach the designated ending point. Defining the number of time points (nodes) to $N$ and a maximum amplitude (angular frequency) of the MW field to $\omega_S^{MW,max}$, one may iteratively determine the  time ($t_k$) and angular ($\theta_k$)  points ($k=1,\dots,N$) by looping $k$ from 2 to $N$-1 (with the initial and end point constrained to $t_1$ = 0, $\theta_1$ = 0 and $t_N$ = $t_m$, $\theta_N$ = $\omega_{\text{eff}}^{(S)}t_m$) as
\begin{eqnarray}
\Delta t_k &=& \xi_t (t_m-t_{k-1}) \frac{\chi}{N-k+2} \quad ,  \label{Eq:29} \\
t_k &=& t_{k-1} + \Delta t_k \quad .\label{Eq:30}
  \end{eqnarray}
Here $\xi_t \in [0,1]$ represents a random number and $\chi$ is selected to limit the time appropriately steps (e.g., $\chi$ = 3), and angular points
\begin{eqnarray}
 \theta_k^{min} &=& \max \{\theta_{k-1}-\omega_S^{MW,max}\Delta t_k, \quad \theta_N-\omega_S^{MW,max} (t_m-t_k)\} \quad ,
  \label{Eq:31} \\
 \theta_k^{max} &=& \min \{\theta_{k-1}+\omega_S^{MW,max}\Delta t_k, \quad \theta_N+\omega_S^{MW,max} (t_m-t_k)\} \quad ,
  \label{Eq:32}
  \\
\theta_k &=& \theta_k^{min} + \xi_a (\theta_k^{max}-\theta_k^{min}) \quad .
  \label{Eq:33}
  \end{eqnarray}
with $\xi_a \in [0,1]$ being a random number. Implemented in a numerical algorithm, the sequential steps outlined in Eqs. (\ref{Eq:29})-(\ref{Eq:33}) enable extremely fast generation of cRW recoupling sequences. These may be implemented directly on the spectrometer, or interpolated to provide equal pulse lengths in the pulse sequence element to reduce phase transients for very short pulse durations.

\begin{figure}
    \centering
    \includegraphics[width=0.75\textwidth]{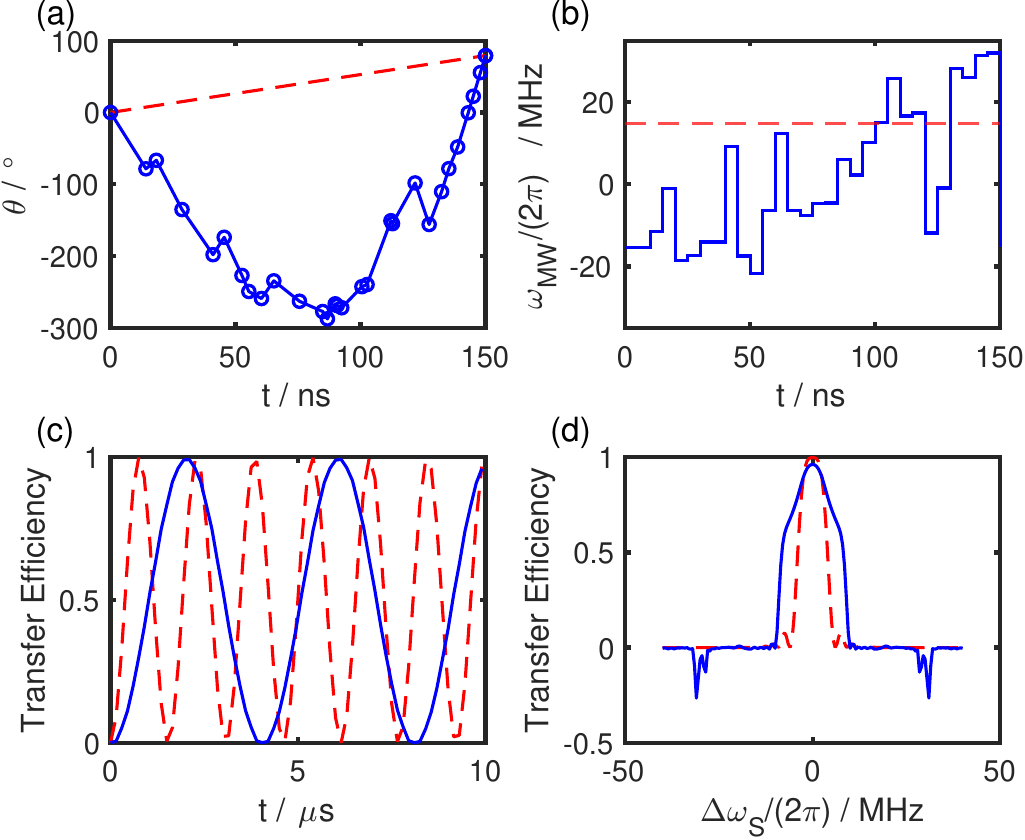}
    \caption{Constrained random walk (cRW) design of DNP experiments in an e$^-$-$^1$H two-spin system (hyperfine coupling constant of $T/(2\pi)$=0.8676 MHz, 45$^\circ$ angle between the e$^-$-$^1$H axis and the external field) subjected to MW irradiation on the electron spin. a) cRW angle trajectory (blue line, time points marked by circles), obtained using 30 pulses  with constrained random amplitude and duration within the limits of a total duration of the element set to $t_m$ = 150 ns, an upper MW amplitude of 32 MHz, and  the desired accumulated end-point angle (end-point of the linear trajectory, upper right corner). The dashed red line corresponds to the rotation angle of NOVEL (14.8 MHz MW amplitude) subtracted $2 \omega_m t_m$.  b) MW pulse sequence ($\pm$x phase, blue line) corresponding to the cRW trajectory in a) with the pulse lengths adjusted to 5 ns steps by interpolation. Build-up c) and MW offset profile d) of the cRW pulse sequence in d) (blue) and the corresponding constant amplitude ($\omega_{\text{MW}}/2\pi$ = 14.8 MHz NOVEL experiment (red dashed) for $S_x$ to $-I_z$ DNP.}
    \label{fig2}
\end{figure}

As a first demonstration of the principle of cRW experiment design, we consider a DNP experiment for an electron-nuclear (e$^-$- $^1$H) spin system influenced by the non-secular hyperfine coupling interaction, an external magnetic field corresponding to $^1$H nuclear Larmor frequency of $\omega_{0I}/(2\pi$) = 14.8 MHz (corresponding to $\omega_{0S}/(2\pi$) = 9.742 GHz for the electron spins, typically referred to as X-band), and MW irradiation on the electron spins (no RF irradiation on the nuclear spin). The hyperfine coupling constant is $T/(2\pi)$ = 0.8676 MHz (corresponding to an electron-nuclear distance of 4.5 \AA) which for a single crystallite with angle $\beta_{PL}$ = 45$^\circ$ between the electron-nuclear spin axis and the external magnetic field corresponds to a pseudo-secular coupling of $B/(2\pi) = \frac{3}{2}T \sin(2\beta_{PL})/(2\pi)$ = 1.3015 MHz (we note that in the frame of the effective MW field, the bilinear terms are proportional to $B/2$). Setting the length of the pulse sequence element of $t_m$ = 150 ns, the number of pulses (nodes) to $N$=30, and the maximum MW amplitude to $\omega_S^{MW,max}/(2\pi)$ = 32 MHz, the cRW method may lead to $S_x$ to $-I_z$ (in the DQ case) polarization transfer pulse sequences of the type illustrated in Fig. \ref{fig1}. We will in the next section argue more for the chosen parameters as part of a more general design strategy. 

Figure \ref{fig2}a shows in red (dashed line) the linear trajectory between the initial and final effective rotation angle, characterized by a linear field matching the $^1$H nuclear Larmor frequency. This angle is given by $(\omega_S^{\text{MW}}-k_I \omega_m) t_m$, where $k_I=2$, corresponding to a subtraction of 2 rotations from the pulse modulation, which in degrees corresponds to a rotation angle $\theta$ = 79.2$^\circ$. The rotation angles taken up by the individual cRW random pulses are marked by a blue line with the circles representing random nodes. The latter, starting and ending at the same angles as the linear trajectory, is generated by a sequence with $x$-phase MW irradiation with amplitude given in Fig. \ref{fig2}b after adjusting (by interpolation) all pulse durations to 5 ns (in Fig. \ref{fig2}a all pulses were allowed to have any duration with the limitation that all 30 pulses should be within $t_m$ = 150 ns duration of the pulse sequence element). We note that this adjustment is not strictly necessary but may ease instrumental implementation of the MW pulse sequence, and potentially reduce effects from phase transients which may be more severe for very short pulses with a duration lower than the rise-fall-time of pulses determined by the Q factor of the resonator. Figure \ref{fig2}c shows buildup curves for the cRW pulse sequence in  Fig. \ref{fig2}b and the constant amplitude NOVEL sequence represented by solid blue and dashed red lines, respectively, with the time axis expanded by concatenating basic building blocks each of length 150 ns up to 10 $\mu s$. Figure  \ref{fig2}d gives the MW offset profile for the two sequences. From this analysis, it evident that cRW may be used to design efficient DNP experiments in a very simple fashion. The arbitrarily selected DNP experiment based on a 30-pulse element provides 100 \% transfer for a single crystal. It has a smaller dipolar scaling factor than the maximum value as provided by NOVEL (i.e., longer time for optimum transfer, Fig. \ref{fig1}), but is somewhat more broadbanded than the NOVEL experiment. 

\section{RESULTS AND DISCUSSION}

While the previous section presented basic elements, we will in this section demonstrate how they can  be combined in a powerful design strategy for systematic development of recoupling experiments and how the FOM-based non-linear optimization procedure as claimed clean up the effective Hamiltonians to provide broadbrand recoupling. We take DNP as the case, noting that the design strategy is equally applicable for heteronuclear dipolar recoupling in MAS solid-state NMR spectroscopy.\cite{DNP_SSNMR_Carvalho:2025aa}

\subsection{Selecting the resonance condition}

 As addressed above, based on Eq. (\ref{Eq:25}), its graphical representation in Fig. \ref{fig2}, and the points discussed in relation to this, the first step is to select an appropriate ZQ or DQ resonance condition which match the present hyperfine coupling interaction and instrumental MW capabilities in terms of the static field and target nucleus (determines the nuclear Larmor frequency; note that the plot in Fig. \ref{fig2} represent 14.8 MHz and that the plot differs depending on $\omega_{0I}$), the available MW field strength (peak power and duty cycle), and potential resonator MW field inhomogeneity. It is important to consider  that the effective field of the $I$ spin $\omega_{\text{eff}}^{(I)}$ (here considered as the nuclear Larmor frequency $\omega_{0I}$ subtracted $k_I \omega_m$, i.e., the first two terms in Eq. (\ref{Eq:25})) and the effective field of the $S$-spin induced by the MW pulse sequence should exceed the effective hyperfine coupling constant (i.e.,  $\omega_{\text{eff}}^{(I)} \approx  \pm \omega_{\text{eff}}^{(S)} > B$) to ensure truncation of effective Hamiltonian components containing only one transverse operator in the frame with the effective fields aligned along $z$. This ensures that the evolution of the density operator in proximity of a ZQ or DQ resonance to first order occurs in ZQ or DQ 3D invariant subspaces and thereby that polarization transfer may be adequately described by the FOM equation in Eq. (\ref{Eq:10}). At the same time, the $S$-spin effective field should be as small as possible to avoid excessive susceptibility of the pulse sequence towards MW field inhomogeneity. 
 
 In the present case with a hyperfine coupling in the order of 1 MHz, we recommend an effective field in the order of 1.5 MHz exceeding the effective hyperfine coupling by a factor of $\approx$3. The other steering parameter for the resonances is the modulation time (i.e., length of the pulse sequence element), which should be selected to match a DQ (or ZQ) resonance condition not in proximity of another ZQ (or DQ) resonance to avoid destructive overlap in the DNP polarization transfer. Along with this comes the choice of $k_I$. A good choice, in this case, will be $t_m$ = 150 ns (corresponding to a modulation frequency of $\omega_m/(2\pi)$=6.667 MHz) and $k_I$=2 corresponding to the DQ resonance marked by dotted lines and a circle in Fig. \ref{fig2}. We note that this choice locks the effective field to $\omega_{\text{eff}}^{(S)}$=1.467 MHz in the optimization process, implying the sequences under consideration does not drift towards effective fields not compatible with the first-order approximation nor sequences excessively susceptible to MW inhomogeneity. While the latter obviously (as demonstrated below) can be further coped with in a subsequent non-linear optimization step in the design process, it proves important  to minimize such effects already at the point of resonance selection. We note that it also possible to address MW inhomogeneity by allowing the pulse sequence in an optimization step to search for sequences with longer total length $t_M$ to allow for adiabatic passage. This aspect will not be addressed further in this work.

\subsection{DNP sequences by constrained random walk }

A second step in the design strategy is identifying good DNP pulse sequences by constrained random walk (cRW), either stand-alone with the aim of random design or as a means to establish good starting points for subsequent non-linear/optimal control optimization. We note that sequences in this case are not optimized, but they can be generated extremely fast  with the best sequences selected among many possible solutions, e.g., by favoring excitation over a large electron spin offset bandwidth to cover radicals with broad EPR lines. 

\begin{figure}
    \centering
    \includegraphics[width=\textwidth]{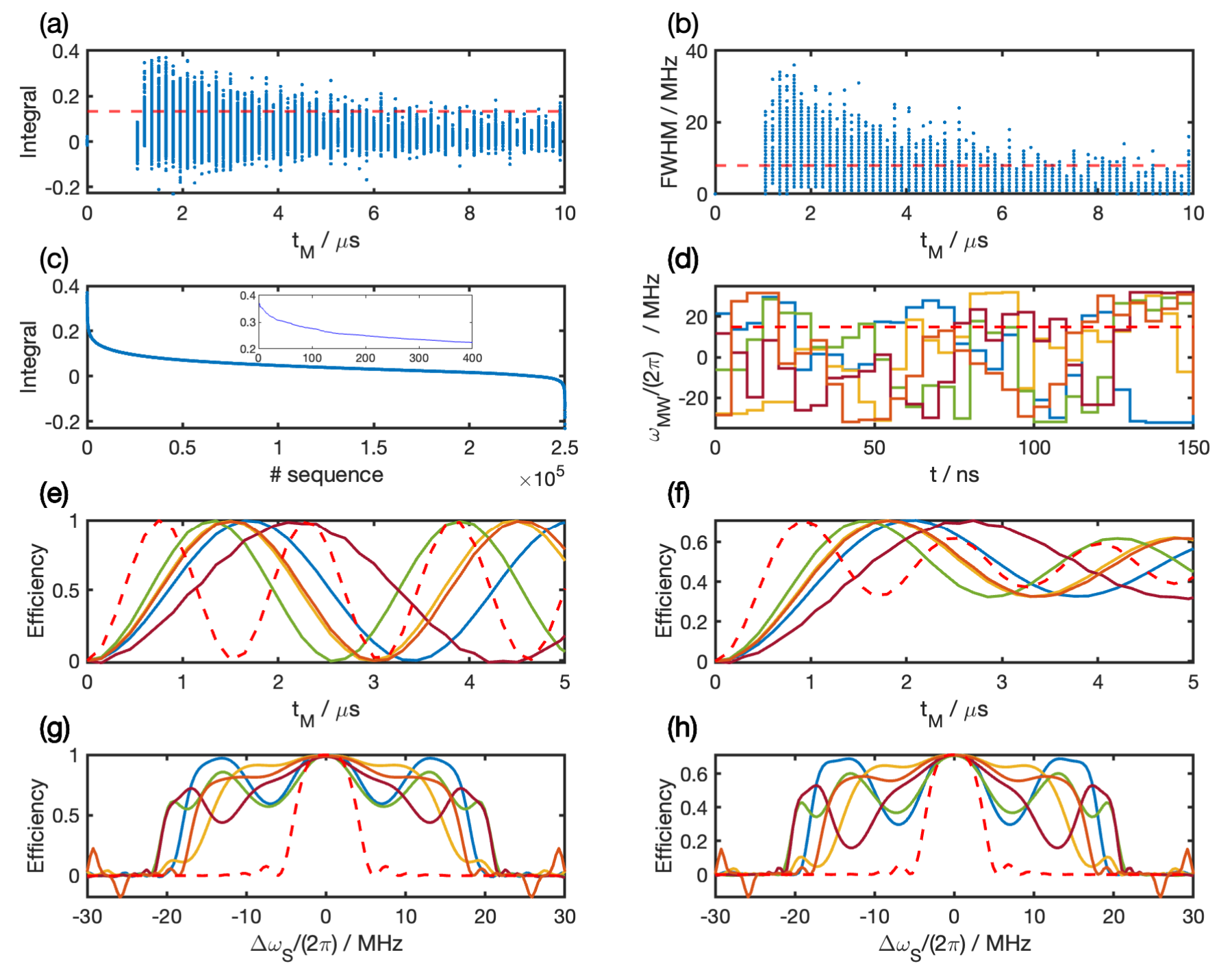}
    \caption{DNP pulse sequences derived using the cRW procedure in Eqs. (\ref{Eq:27})-(\ref{Eq:33}). 250,000 cRW sequences were calculated using $t_m$ = 150 ns with $N$ = 30 $\pm x$-phase pulses (MW amplitude limited to 32 MHz) and duration altogether reaching the resonance condition in Eq. (\ref{Eq:27}) with $k_I$=2 and $\omega_{0I}/(2\pi)$ = 14.8 MHz (same field and hyperfine coupling parameters as used in Fig. \ref{fig2}). The sequences were interpolated to provide elements with 5 ns pulses, and repeated to a total MW sequence length $t_M=n t_m$ to establish maximum transfer in single crystal density operator calculations ($n$ is sequence-specific as it depends on the scaling of the pseudo-secular hyperfine coupling by the MW pulse sequence). (a,b) The 250,000 cRW sequences characterized by the integrated intensity over $\pm$ 20 MHz offset range (a) and FWHM of offset profile. (c). The sequences in (a) sorted with the highest integral first for all 250,000 sequences and 400 sequences in the insert. (d-h) The 5 best pulse sequences from (c) represented with (d) pulse sequence element, (e,f) single crystal (e) and powder (f) buildup curves, and (g,h) single crystal (g) and powder (h) MW offset profiles. The curves in (e-h) were obtained by density operator calculations with $\rho(0)=S_x$ and detection by projection onto $-I_z$. In (a), (b), (d) - (h) the dashed red line corresponds to NOVEL (constant x-phase MW irradiation with amplitude $\omega_S^{\text{MW}}/(2 \pi)$ = 14.8 MHz).}
    \label{fig3}
\end{figure}

To demonstrate these aspects, we select according to the criteria above (and as also used for the initial demonstration of cRW in Fig. \ref{fig2}) sequences with a pulse sequence element of duration $t_m$ = 150 ns. We select $N$ = 30 pulses which in a subsequent step may be coped with using standard non-linear optimization provided only the MW amplitude is varied (i.e., pulses of phase $\pm$x). The modest number of pulses may also (upon interpolation, see below) give relatively long pulses (5 ns) potentially being less susceptible to phase transients than sequences also including much shorter pulses. These choices obviously depend on the mode of non-linear/optimal control optimization, including parameters used in the optimization, on the targeted spin system, and on the instrumentation on which the resulting pulse sequences should be implemented. 
 Using the described cRW procedure, we calculated 250,000 pulse sequences and analyzed these by density operator calculations with the elements repeated to reach the  maximum transfer for the chosen single crystal orientation and with this powder distribution of crystallite orientations. The resulting cRW sequences may readily be characterized in terms of broadbandedness as quantified as either the integral of the excited signal within $\pm$20 MHz (40 MHz bandwidth) or as the full width at half maximum (FWHM) as illustrated in Figs. \ref{fig3}a and \ref{fig3}b, respectively. In Fig. \ref{fig3}c the 250,000 sequences are sorted according to integrated transfer efficiency over a 40 MHz bandwidth, and the best sequences taken out for discussion and further optimization. 
 
To facilitate the analysis, we here illustrate the most important aspects by characterizing the five best pulse sequences (interpolated to give 5 ns pulse lengths) as shown in Fig. \ref{fig3}d. For each of these pulse sequences, Figs. \ref{fig3}e - \ref{fig3}h shows density operator build-up curves (Figs. \ref{fig3}e, \ref{fig3}f) and offset profiles (Figs. \ref{fig3}g, \ref{fig3}h) for the single crystal used in the cRW calculation (Figs. \ref{fig3}e, \ref{fig3}g) and a powder with averaging over the angle $\beta_{PL}$ discriminating the electron-nuclear axis and the direction of the external field (Figs. \ref{fig3}f, \ref{fig3}h). We note that the single-crystal and powder representations apart from an overall scaling factor are largely identical enabling design on basis of a single crystal orientation only.\cite{DNP_SSNMR_Carvalho:2025aa} For comparison Figs. \ref{fig3}e - \ref{fig3}h also include the corresponding calculations for the NOVEL experiment marked with dashed red lines. From this presentation, it is clear that cRW can derive pulse sequences with much larger broadbandedness than NOVEL, and thereby in itself represents a valuable stand-alone tool for pulse sequence design. The best sequences cover a bandwidth of 30-40 MHz being 3-4 times that of NOVEL and almost reach the performance of the recent BEAM method.\cite{BEAM} It is, however, also clear that the excitation profiles are not completely uniform over the relevant offset range, which may - along with increasing bandwidth - be an objective for subsequent numerical optimization as demonstrated in the next section.

\subsection{FOM-based effective Hamiltonian optimization of broadband DNP experiments}

 With good starting guesses at hand, non-linear or optimal control procedures can be used to further improve the pulse sequences with our focus here being broadbandness toward electron spin offsets, as was also the target for the cRW initial guesses. The basis for the optimization is the FOM fidelity function in Eq. (\ref{Eq:10}) where both the bilinear terms (defined by $\omega_{\text{bil}}$) driving polarization transfer and the non-commuting linear terms (defined by $\omega_{\text{lin}}$) deteriorating the transfer (if not stabilized) are optimized simultaneously. Exact effective Hamiltonian theory (EEHT)\cite{EEHT} in SU(4), as formulated in Eqs. (\ref{Eq:17}) - (\ref{Eq:24}), is used to derive the underlying single- and two-spin frequency components as specified in Eqs. (\ref{Eq:11}) - (\ref{Eq:16}) in a frame with the effective fields for the two spin species along the longitudinal axis. Definition of the effective field frame is trivial in this case for the nuclear spins as no RF irradiation is invoked, while for the electron spin, it found using EEHT in the SU(2) formulation (cf., Eq. (\ref{Eq:9})). 
 
 For the linear and bilinear fields, the effective Hamiltonian and its projections onto the one- and two-spin operators as defined in Eqs. (\ref{Eq:15}) and (\ref{Eq:16}), respectively, need to be calculated only for the pulse sequence element. Projection into the future by repetition of these elements is taken into account via the parameter $t_M$ in Eq. (\ref{Eq:10}). This "forward prediction" provides a tremendous reduction in the computational effort in numerical optimizations relative to pulse sequences with pulses varying throughout the full period $t_M$ (as typical for state-to-state optimal control), and makes it much easier to design pulse sequences which can be adapted to differently sized hyperfine couplings through adjustment of the number of repetitions of the basic element. The overall length of the recoupling sequence, being expressed as the number $n$ of repetitions to obtain a mixing period of $t_M=nt_m$, is most conveniently  defined from cRW starting guesses to pair the initial MW sequence with its pseudo-secular hyperfine coupling scaling factor. We note that starting guesses may obviously be selected/filtered both according to broadband performance and buildup efficiency (scaling factor) as fast buildup may be advantageous in the presence of dissipation due to fast relaxation.

\begin{figure}
    \centering
    \includegraphics[width=\textwidth]{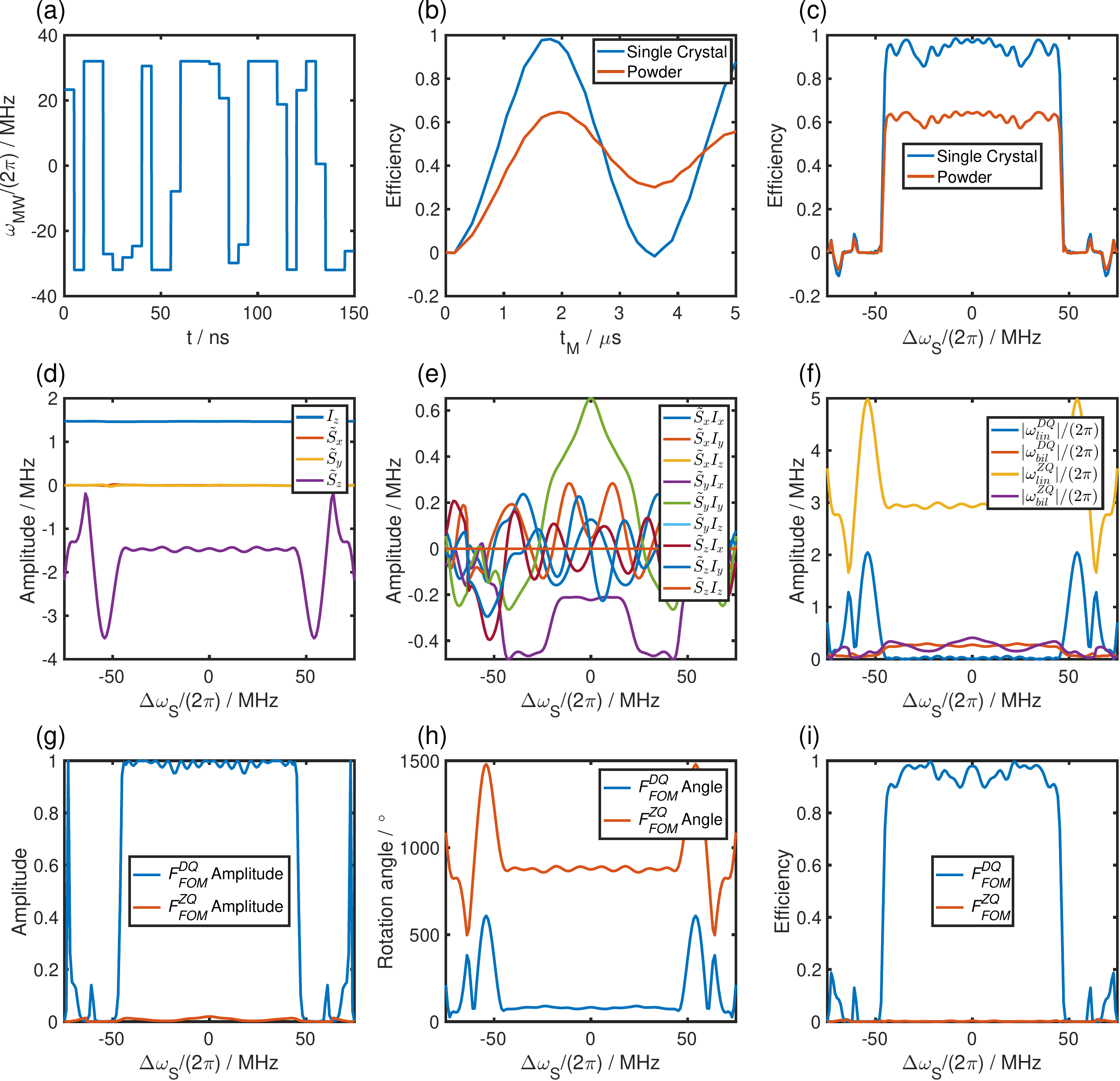}
    \caption{DNP pulse sequence optimized using FOM-based non-linear optimization. The sequence resulted from optimization of the best cRW pulse sequences in Fig. \ref{fig3} with $t_m$ = 150 ns 30-pulse sequence elements with  $t_M$ fixed to the value coming out of the selected cRW pulse sequences and optimized to provide an electron-spin bandwidth (offset range) of 100 MHz. The optimization was performed without consideration of MW inhomogeneity, rationalized by selection of the DQ resonance with $k_I$=2 and an electron spin (S) effective field of $\omega^{(S)}_{\text{eff}}/(2\pi)$ = -1.467 MHz. (a) Pulse sequence element, (b) density operator build-up curve for the single crystal used for optimization and a powder, and (c) the corresponding offset profile of the pulse sequence for single crystal and powder conditions. (d-i) Illustration of determinants of the FOM-based, EEHT-assisted effective Hamiltonian optimization by offset plots of (d) linear fields, (e) bilinear fields, (f) linear and bilinear frequencies, (g) FOM amplitude, (h) FOM angle, and (i) FOM function.}
    \label{fig4}
\end{figure}

While optimal control procedures as recently presented in Ref. \cite{OC_DNP} can be used, we use here simple Nelder-Mead type SIMPLEX methods\cite{simplex,press2007numerical} as implemented in Matlab \cite{MATLAB:2010} to optimize the FOM fidelity $\mathcal{F}_{\text{FOM}}^{DQ}(t_M)$ function for a single crystallite (parameters given above) with the target being 100 MHz electron-spin offset bandwidth and under constraints of a maximum MW amplitude of 32 MHz. With intrinsic low susceptibility to MW inhomogeneity effects through a deliberately low $S$-spin effective field, the initial optimization was accomplished assuming homogeneous MW field conditions. The optimization lead to the 30-pulse sequence illustrated in Fig. \ref{fig4}a characterized by the density-operator calculated build-up and offset profiles shown in Figs. \ref{fig4}b and \ref{fig4}c, respectively, with blue and red curves representing single-crystal and powder cases, respectively. It is seen that the pulse sequence resulting from optimization increases the bandwidth of approximately 40 MHz from the best cRW initial guesses to 100 MHz, and markedly improves the uniformity of excitation of the desired offset range. It is noted that we optimized pulse sequences from multiple cRW starting guesses, generally leading to well-performing broadband DNP pulse sequences. Out of these, we present four sequences in the experimental section together with one control sequence fully optimized with the replicated state-to-state method proposed in Ref. \cite{OC_DNP} employed directly on cRW starting guesses. For the detailed analytical analysis presented in this section, we consider only one of these sequences optimized using the Nelder-Mead type SIMPLEX methods.

Appreciation of the determinants of an optimal pulse sequence, and thereby the power of (and justification of premises for) FOM-based optimization as a means to optimize the relevant effective Hamiltonian, may be provided by Figs. \ref{fig4}d - \ref{fig4}i setting focus on the various components of  $\mathcal{F}_{\text{FOM}}^{DQ}(t_M)$ as function of the electron-spin offset. Figure \ref{fig4}d illustrates the trajectories of the linear fields over the offset range. As requested by selecting the resonance marked by the cross in Fig. \ref{fig1}, the $I_z$ trajectory is defined uniquely by an $^1$H effective field of $\omega^{(I)}_{\text{eff}}/(2\pi)$=$(\omega_{0I}-2\omega_m)/(2\pi)$ = 1.467 MHz using a nuclear Larmor frequency of 14.8 MHz, a pulse sequence modulation frequency of 6.667 MHz, and by choosing the $k_I$=2 resonance. No other $I$-spin components can contribute as the nuclear spins are not subjected to RF irradiation. The transverse $\tilde{S}$ spin fields are zero, as they should be in a frame with the effective MW field along the longitudinal axis. The $\tilde{S}_z$ profile is flat with the desired resonance matching effective field of $\omega_S^{\text{eff}}/(2\pi)$=-1.467 MHz over the desired bandwidth. 

The projections onto the bilinear operators, as represented in Fig. \ref{fig4}e, are not equally straightforward to interpret as the optimization does not optimize any of these specifically. These are optimized through the norm of the $x-$ and $y-$phase bilinear terms as formulated in Eq. (\ref{Eq:14}). The functionality of these optimized norm terms are given in Fig. \ref{fig4}f, where it can be seen that the linear DQ-subspace term (proportional to $\frac{1}{2}(\tilde{S}_z+I_z)$) has a very small amplitude ($<$ 60 kHz) over the targeted $S$-spin band width, while the corresponding ZQ-subspace term (proportional to $\frac{1}{2}(\tilde{S}_z-I_z)$) is much larger (in the order of 2.95 MHz) which implies that the bilinear DQ terms (amplitude in order of 0.3 MHz) are not truncated by the linear fields, while the bilinear ZQ terms (similar magnitude) are truncated by the large linear ZQ term being substantially larger than the planar ZQ-subspace terms (proportional to and smaller than the maximal effective pseudo-secular hyperfine coupling of $\frac{1}{2}B/(2\pi) \approx$ 0.65 MHz (for $\beta_{PL}=\pi/4$) used in the optimization. The linear and bilinear terms enter directly into the amplitude of the FOM function ($ |\omega_{\text{bil}}^p|^2/(|\omega_{\text{bil}}^p|^2+4|\omega_{\text{lin}}^p|^2)$) and the scaling factor in the $\sin^2$ function ($\frac{t_M}{4}\sqrt{|\omega_{\text{bil}}^p|^2+4|\omega_{\text{lin}}^p|^2}$) as mapped in Figs.  \ref{fig4}g and \ref{fig4}h, respectively. The amplitude factor clearly reveals possible excitation only through the DQ terms and the contours of the ideal excitation function, while the latter  through angles stabilized in proximity of 90 $^\circ$ and 880$^\circ$ for the DQ and ZQ, respectively, also support selection of the DQ resonance. All this accumulates to the FOM function in Fig. \ref{fig4}i, with a profile closely matching the result of the density operator calculations in Fig. \ref{fig4}c for the single crystal, and after a scaling also for the powder. This was the aim of the optimization: agreement between the FOM and density operator profiles with effective Hamiltonian control, which supports fulfillment of the first-order approximation of 3D DQ subspace evolution and, with this, the validity of the fidelity function  $\mathcal{F}_{\text{FOM}}^{DQ}(t_M)$ laid out in Eq. (\ref{Eq:10}).

\subsection{Time-dependent effective Hamiltonians through auto-truncation}

An important new element in our design strategy is to develop pulse sequences - that be DNP or any other type of magnetic resonance pulse sequences - which intrinsically auto-truncate (or self-truncate) the effective Hamiltonian through repetition. This at first sight peculiar example of time-dependent effective Hamiltonians, supplements the over decades widely used strategy of concatenating phase-modulated pulse sequences elements in the design of, e.g., composite pulses,\cite{LEVITT_compositepulses} decoupling,\cite{LEVITT198347} and solid-state NMR homonuclear\cite{scBCH,Rhim3,BURUM1981173,MSHOT3,UHDS,MOTE20161} and heteronuclear decoupling\cite{Eden_decoupling} and recoupling\cite{C7,POSTC7,Levitt_symm_rec} solid-state NMR pulse sequences. 

\begin{figure}
    \centering
    \includegraphics[width=\textwidth]{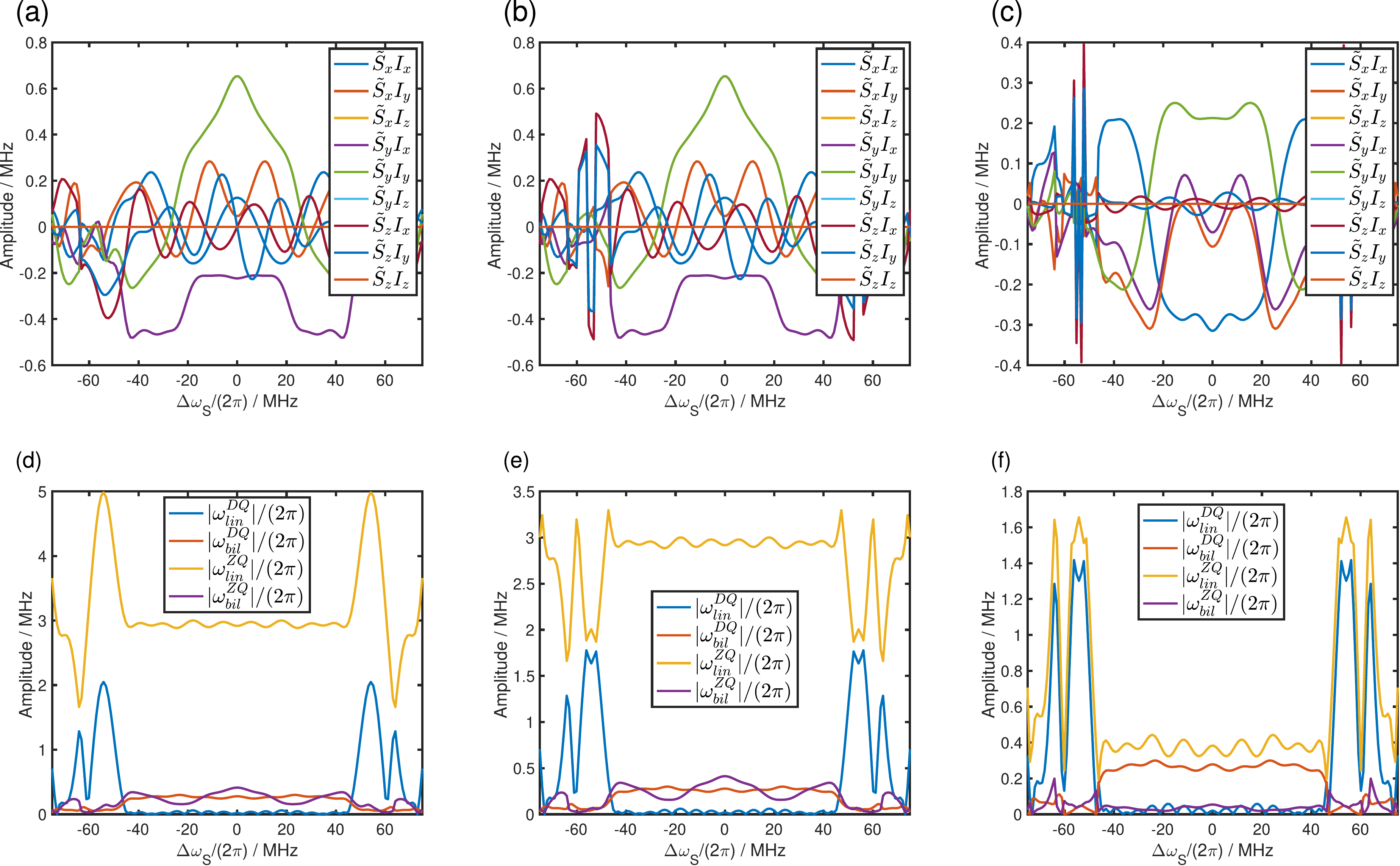}
    \caption{Illustration of auto-truncation of the effective Hamiltonian for the broadband DNP pulse sequence represented in Fig. \ref{fig4} to provide basis for pure DQ-subspace dipolar recoupling.  The figure reproduces the plots in Figs. \ref{fig4}e and  \ref{fig4}f just with the element repeated $N$=1 (a,d), $N$=2 (b,e), and $N$=4 (c,f) times. (a-c) Individual bilinear components of the effective Hamiltonian and (d-f) linear and bilinear components of the FOM function in Eq. (\ref{Eq:10}) calculated using EEHT with the formula in Eqs. (\ref{Eq:11}) -  (\ref{Eq:24}).}
    \label{fig4b}
\end{figure}

Following up on the introduction to the concept of auto-truncation of the internal Hamiltonian in the theory section, Figure \ref{fig4b} illustrates this through EEHT-based calculation of relevant parts of the effective Hamiltonians of the DNP pulse sequence in Fig. \ref{fig4} upon 1, 2, and 4 repetitions. Figures \ref{fig4b}a and \ref{fig4b}d replicate (for comparison) the plots in Figs. \ref{fig4}e and  \ref{fig4}f with bilinear (two-spin) effective Hamiltonians and the size of linear and bilinear components of the FOM function, respectively. From Figs. \ref{fig4b}b,\ref{fig4b}e and \ref{fig4b}c,\ref{fig4b}f showing the same trajectories but for $N$=2 and $N$=4 repetitions, respectively, it becomes clear ($i$) that two repetitions do not alter significantly the effective Hamiltonian while ($ii$) four repetitions markedly changes both the individual components, as well as the linear and bilinear ZO/DO subspace components being ingredients in the FOM fidelity function in Eq. (\ref{Eq:10}). Comparing first the linear and bilinear fields in Fig. \ref{fig4b}e with those in Fig. \ref{fig4b}c, it is clearly visible that the bilinear ZQ component present in Fig. \ref{fig4b}c is not present to near the same extent in Fig. \ref{fig4b}d. This is ascribed to the auto-truncating effect of the large linear ZQ field upon repetition of the element followed by evaluation of the effective Hamiltonian. In contrast, the corresponding bilinear DQ operators remain as they are not subject to truncation by a non-present (i.e., much smaller) linear DQ field. This effect also becomes visible in the individual bilinear Hamiltonian components, as seen by comparing Figs.  \ref{fig4b}a and \ref{fig4b}c. It is here evident that a $DQ_x= \tilde{S}_xI_x-\tilde{S}_yI_y$  effective Hamiltonian component is present largely for all offsets, although changing sign and vanishing around $\pm$ 25 MHz offset, but here replaced by a large $DQ_y= \tilde{S}_yI_x+\tilde{S}_xI_y$ component. Both of these Hamiltonians contribute to the $|\omega_{\text{bil}}^{ZQ}|$ component of Eq. (\ref{Eq:12}) to the FOM function, thereby forming the fundament for efficient 2Q DNP transfer.

\subsection{Refinement to cope with inhomogeneous MW fields}

\begin{figure}
    \centering
    \includegraphics[width=0.75\textwidth]{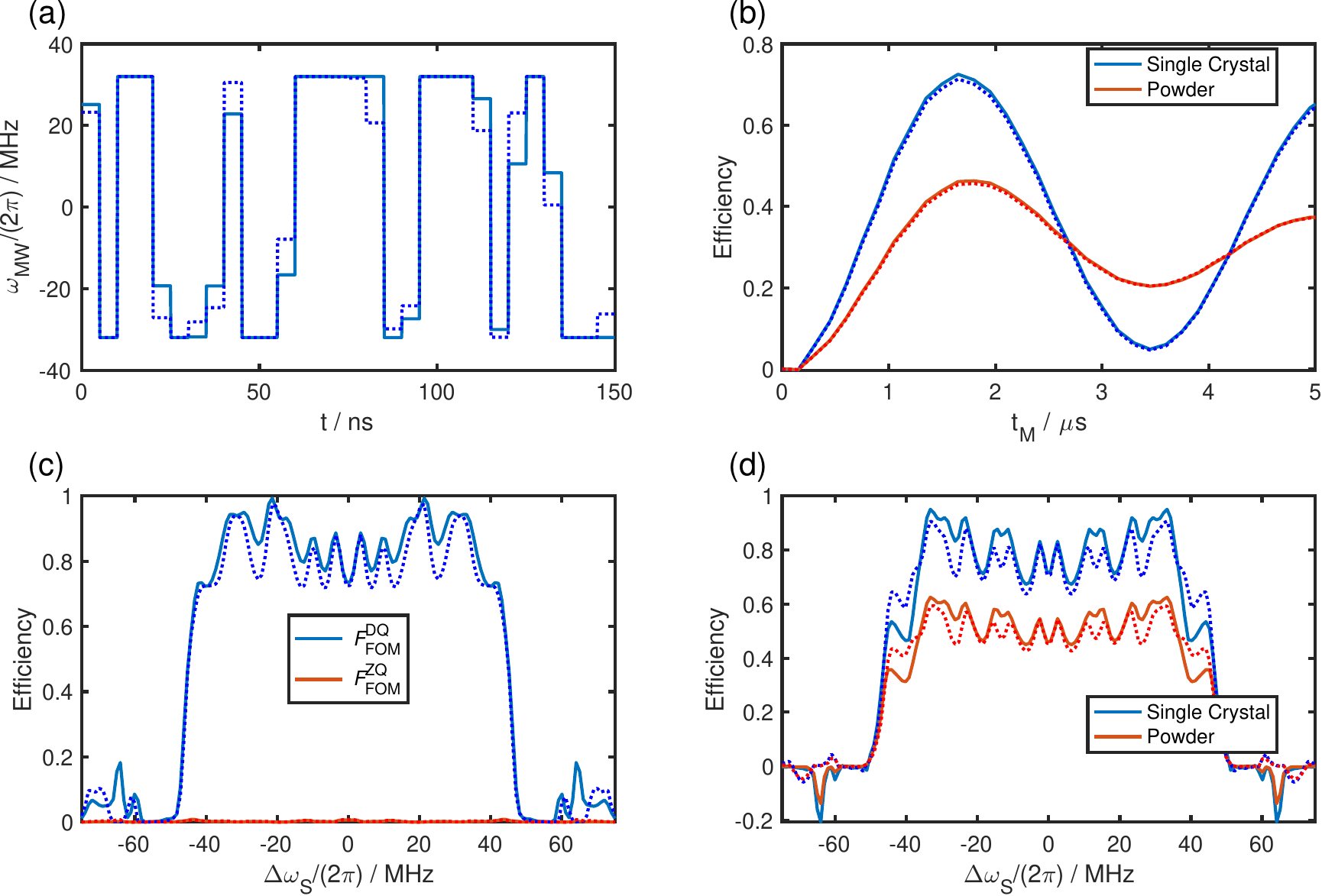}
    \caption{MW inhomogeneity compensated cRW-OPT variant of the pulse sequence in Fig. \ref{fig4}a - obtained after reoptimizing with an experimentally determined MW inhomogeneity profile (see Experimental). (a) cRW-OPT pulse sequence, (b) density operator contact-time build-up curve, (c) FOM fidelity function (cf. Eq. (\ref{Eq:10})), and (d) density operator $S$-spin offset profile. In (b) and (d) the blue curve represent a single crystal while the red curve represent a powder. The initial sequence (Fig. \ref{fig4}) is represented by dotted lines.}
    \label{fig5}
\end{figure}

Aimed at optimizing the pulse sequence optimally to the MW inhomogeneity determined experimentally (see the profile in the Experimental Section) for our X-band pulsed DNP spectrometer,  the pulse sequence in Fig. \ref{fig4}a was reoptimized under consideration of MW inhomogeneity to give the cRW-OPT pulse sequences in Fig. \ref{fig5}a with the corresponding build-curve (Fig. \ref{fig5}b), FOM function offset dependency (Fig. \ref{fig5}c), and density operator offset profile (Fig. \ref{fig5}d). While it is evident that the pulse sequence only changed slightly (the characteristics of the original sequence under inhomogeneous MW field conditions is represented by dotted lines), the resulting cRW-OPT pulse sequence compensates very well for the inhomogeneity in terms of broadbandedness and overall polarization transfer despite not giving as ideal square offset profiles as in the homogeneous MW field  case. The 100 MHz bandwidth is maintained with around 60\% transfer efficiency over this range for the powder cases. It is worth mentioning the remarkable aspect that the original sequence due to the deliberately chosen low $S$-spin effective field within small margins can not be improved significantly by reoptimization. The overall fidelity increased only from 0.78 to 0.83 in the present case. This implies that optimal experiments not only can be developed on basis of a single crystal orientation, but also for a single MW field isochromat upon reasonable prior consideration of the optimization problem. In combination, these aspects reduce the effort of numerical optimization enormously.

\subsection{Experimental demonstrations}

\begin{figure}
    \centering
    \includegraphics[width=\textwidth]{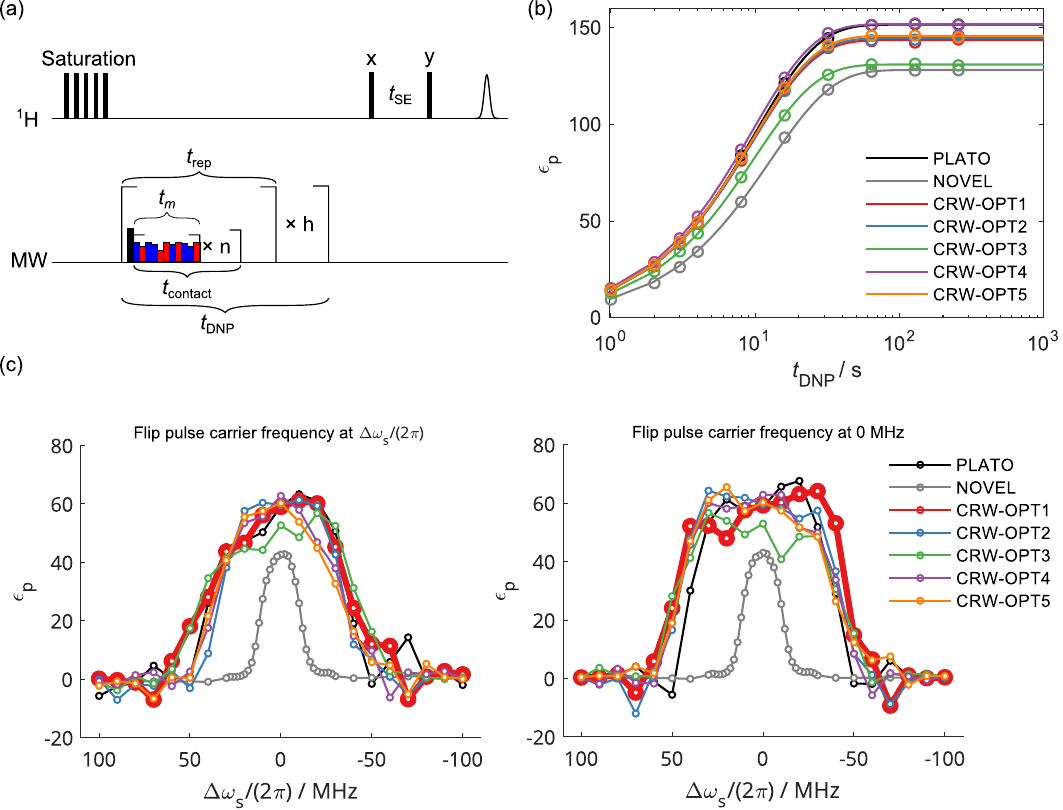}
    \caption{(a) MW (e$^-$) and RF ($^1$H) pulse scheme used in experimental implementation of cRW-based non-linear optimization e$^-$-$^1$H DNP pulse sequences. The pulse scheme includes initial saturation pulses and readout via a solid-echo at the $^1$H RF channel in addition to the $(\pi/2)_y$ initial pulse and the DNP pulse sequence at the electron-spin MW channel. (b-c) Experimental data for for obtained for OX063 at 80 K using a home-built pulsed DNP/EPR spectrometer operated at X-band with the pulse sequence in (a) comparing the performance of PLATO, NOVEL, and five cRW-OPT pulse sequences (see parameters in Experimental). (b) $^1$H enhancement factors ($\epsilon_{\text{p}}$) as a function of pumping time ($T_{\text{DNP}}= h\, t_{\text{rep}})$ at $\Delta \omega_S/(2\pi)=0$ MHz. All buildup times, $T_{\text{B}}$, defined through $\epsilon_p(T_{\text{DNP}}) =
\epsilon_{\text{max}}[1 - \exp(-T_{\text{DNP}}/T_{\text{B}})]$ are in the order of 10 s (\textit{c.f.} Tab. \ref{tab1})
(c)  $^1$H signal enhancements measured as function of the electron-spin offset $\Delta \omega_S/(2\pi)$ under assumption of the initial electron-spin $\pi/2$ pulse being moving with the offset as the DNP pulse sequence element (left) and  on-resonance throughout the offset scan of the DNP sequences (right). }
    \label{fig6}
\end{figure}

With a rational design approach and demonstrated resulting development of high-performance broadband DNP pulse sequences in place remains the important aspect  of illustrating the capabilities of the developed methods experimentally. For this purpose, we used a home-built pulsed EPR/DNP instrument operating at X-band EPR frequencies (details in the Experimental section) with optimized  cRW-OPT DNP pulse sequence elements (as in Figs. \ref{fig5}a) implemented into the general RF/MW pulse sequence framework in Fig. \ref{fig6}a. The DNP pulse sequences were implemented with timings and amplitudes resulting from numerical optimization (see below) with an experimentally optimized number of repetitions, see Table \ref{tab1},  to cope with  larger spin systems. The experimental demonstrations were performed  for an OX063 trityl sample at 80 K. We note that this sample  is characterized by a much narrower EPR line than the electron-spin excitation bandwidth of the proposed DNP sequences. While realizing that  more impressive sensitivity boosts may have been visualized using  samples with broader EPR liner, this particular sample was deliberately chosen to examine and document in detail the offset profiles (i.e., control of the effective Hamiltonian) of the cRW-OPT DNP pulse sequences as being a keypoint in our theoretical and numerical analysis. This includes exploring the influence from the initial electron-spin $(\pi/2)_y$ pulse bringing initial electron-spin ($S_z$) polarization to $S_x$. Further exploration of the performance on other radicals or metal centers is beyond the scope of this paper, and will be subject to future studies.

Prior to implementing the pulse sequences on the spectrometer, we took four pulse sequences optimized using the Nelder-Mead type SIMPLEX method and refined them in the OC-based replicated state-to-state optimization software recently introduced by Carvalho et al,\cite{OC_DNP} also considering MW inhomogeneity, to verify performance and convergence to optimal sequences in an optimal control setup. Overall the sequences converged to very similar pulse profiles, some with convergence in just a few iterations thereby leading to essentially the same pulse sequence, while others converged to pulse sequences with small differences (similar to  the difference between original and final optimizations in Fig. \ref{fig5}) in pulse waveforms, but with very similar performance, thereby confirming the methods developed by the non-linear optimization-based design strategy proposed in this work. This observation is further strengthened by the control sequence fully optimized with replicated state-to-state, also having similar numerical performance. The four selected pulse sequence elements, cRW-OPT(1-3,5) together with the control sequence, cRW-OPT4, are specified in the Experimental Section.

DNP-pumping build-up curves for NOVEL, PLATO, and the five cRW-OPT pulse sequences are shown in Figure \ref{fig6}b. We note that the highest transfer is obtained for some of the cRW-OPT sequences and PLATO, while it is also evident that all sequences  are  characterized by very similar build-up time constants in the order of $T_B \approx$ 10 s. Using a build-up time of $t_{buildup}$ = 5 s, Figures \ref{fig6}c and \ref{fig6}d show  electron-spin offset profiles for the same DNP sequences recorded with fixed (zero) offset for the initial $(\pi/2)_y$ pulse in Fig. \ref{fig6}c and varying (along  with the DNP sequence) offset for this pulse in Fig. \ref{fig6}d. While being almost identical in performance, the pulse sequence (cRW-OPT-1) showing the best offset profile is marked with thicker line. We note that the experimental data in Fig. \ref{fig6}d reflects the situation where the cRW-OPT sequence is used for DNP from unpaired electrons with a broad EPR line, while the implementation in Fig. \ref{fig6}c explores directly the broadbandedness of the cRW-OPT elements as optimized in Figs. \ref{fig4} and \ref{fig5} under assumption of an initial operator of $S_x$. We also note that the two sets of curves are pretty similar and that the best sequences also approach 100 MHz bandwidth in the experimentally most relevant case. 

The offset plots  clearly documents the capability of our effective Hamiltonian approach to systematically design DNP pulse sequences reaching extremely large (targeted 100 MHz) excitation bandwidth with the five chosen sequences being a signature of robustness in the design procedure. The sequences are compared to NOVEL and PLATO pulse sequences, where the former widely used sequences is outperformed significantly by the cRW-OPT pulse sequences in terms of broadbandedeness. The broadbandedness increased from the 80 MHz PLATO sequence to the 100 MHz cRW-OPT DNP sequences as predicted by theory. The plots in Fig. \ref{fig6}c also illustrate that a non-ideal $\pi/2$ initial pulse obviously influences the overall broadbandedness - an effect that beyond this study may be reduced by optimizing composite excitation pulses or targeting $S_z$ to $I_z$ transfer as an alternative to the $S_x$ to $I_z$ transfer here serving to demonstrate our pulsed DNP design strategy.

\section{EXPERIMENTAL}
All experiments were conducted on a home-built X-band pulsed EPR/DNP spectrometer (based on the design of Doll \textit{et al.} \cite{Doll:2017wb}) equipped with a Keysight M8190A 14 bit/8GSa and Z$\ddot{u}$rich Instruments HDWAG4 16 bit/2.4 GSa (Z$\ddot{\text{u}}$rich) Instruments, Z$\ddot{\text{u}}$rich), CH) arbitrary waveform generators on electron and nuclear spins, respectively, a 2 kW Applied Systems Engineering 176 X TWT MW amplifier, a SpinCore  GX-5 iSpin NMR console (SpinCore Technologies Inc., Gainesville, FL), a 300 W NanoNord RF amplifier (NanoNord A/S, Aalborg, Denmark), and a Bruker MD4 electron-nuclear double resonance probe (Bruker BioSpin, Rheinstetten, DE) extended with an external tuning and matching circuit. 

Experiments were performed on a sample of 5 mM trityl (OX063) in a H$_2$O:D$_2$O:Glycerol-d$_5$ solution (1:3:6 by volume) at 80 K (note the higher degree of protonation in this system relative to the commonly used DNP juice; which due to the higher concentration of $^1$H has the consequence of lower overall polarization transfer). Experiments were conducted using the DNP polarization transfer pulse sequence Fig. \ref{fig6}a  which in addition to the proposed cRW-OPT and PLATO\cite{PLATO_adv} DNP  pulse sequences used an initial saturation of 
$^1$H with a set of $S=11$ pulses of duration 1.32 $\mu$s separated by $\tau_{\text{sat}}=1$ ms and a solid-echo sequence  $\pi / 2 - \tau - \pi / 2$ for read out with $\tau$ = 25 $\mu$s; all RF pulses used an  RF field strength of 227.3 kHz corresponding to a $\pi/2$ pulse time of 1.10 $\mu$s. The initial electron spin $(\pi/2)_y$ pulse used an MW field strength of 40.0 MHz, corresponding to a pulse length of 6.25 ns.
 A 5 s overall pumping time with  $P$ = 2500 and $\tau_{\text{rep}}=2$ ms were used for the experiments. Experimental polarization enhancements (denoted $\epsilon_{\text{p}}$) is defined by the ratio between the DNP-enhanced signal intensity and the thermal equilibrium signal intensity. 
\setlength{\tabcolsep}{10pt} 
\begin{table}[h!]
\begin{tabular}{c c c c}
\hline
$t_{\text{contact}}$ (ns) & $\omega_{\text{MW}}/(2\pi)$ (MHz) & $\epsilon_{\text{max}}$ & $T_{\text{B}}$ (s) \\
\hline
2000 & 13.6 & 128.1 & 12.7  \\
840  & 30.0 & 151.6 & 10.0 \\
750  & 32.4 & 143.6 & 9.5  \\
750  & 32.4 & 144.6 & 9.6 \\
600  & 31.6 & 131.0 & 9.9 \\
900  & 32.4 & 151.8 & 9.4 \\
750  & 32.4 & 145.8 & 9.6 \\
\hline
\end{tabular}
\caption{Contact time ($t_{\text{contact}}$), DNP contact MW field strengths ($\omega_{\text{MW}}/(2\pi)$), maximal enhancements ($\epsilon_{\text{max}}$),  and buildup time ($\tau_{\text{buildup}}$ ) for the tested DNP sequences. All sequences were  experimentally optimized individually at $\Delta \omega_S/(2\pi)=0$ MHz. The enhancement factors were also obtained at $\Delta \omega_S/(2\pi)=0$ MHz.}
\label{tab1}
\end{table}

 We compare experimentally five cRW-OPT DNP pulse sequences with NOVEL\cite{NOVEL} and PLATO.\cite{PLATO_adv} The five cRW-OPT DNP pulse sequences are each represented by 30 MW $x$-phase pulses (negative amplitude corresponds to $-x$ phase)  of duration 5 ns. cRW-OPT1: \{-1.918, 32.000, 32.000, 32.000, -5.240, -31.762, -26.469, 29.238, 31.332, 32.000, 32.000, 29.960, -32.000, -31.575, 25.965, 4.065, -32.000, -32.000, -32.000, 15.738, 31.993, -3.823, -32.000, -32.000, -28.383, -26.046, -17.482, 32.000, 27.241, 20.561\}. cRW-OPT2: \{24.930, -31.997, 31.994, 31.999, -19.203, -31.998, -31.862, -19.197, 22.976, -31.987, -31.999, -16.642, 31.999, 31.992, 31.951, 31.939, 31.997, -31.867, -27.317, 31.987, 31.990, 32.000, 26.673, -29.895, 10.726, 31.986, 8.227, -31.998, -31.995, -31.984 \}. cRW-OPT3: \{32.000, 32.000, 1.817, -31.959, 20.542, 32.000, 32.000, 32.000, -31.977, -29.006, 31.963, 32.000, 31.669, 32.000, 32.000, -15.637, -32.000, -26.882, 22.464, -31.532, -29.385, -31.374, 9.479, 32.000, 31.951, -32.000, -32.000, -31.231, -32.000, -7.571 \}. cRW-OPT4: \{1.821, -32.000, 32.000, 32.000, -3.409, -32.000, -32.000, -18.336, 8.941, -32.000, -32.000, -22.110, 23.773, 28.492, 32.000, 32.000, 32.000, -30.780, -31.998, 32.000, 32.000, 32.000, 32.000, -32.000, 23.888, 32.000, 12.127, -32.000, -32.000, -12.671 \}. cRW-OPT5: \{-8.753, -32.000, -32.000, -32.000, 31.131, 30.618, 4.056, -32.000, 32.000, 32.000, 32.000, 29.569, -32.000, -14.760, 32.000, 32.000, 25.391, 31.986, 32.000, -32.000, -32.000, -20.103, 10.126, -29.556, -32.000, -32.000, 8.594, 32.000, 7.516, 2.124 \} MHz. We note that cRW-OPT2 is the sequence analyzed in Fig. \ref{fig5}a (and in preceding homogeneous field form in Fig. \ref{fig4}a.

Numerical optimizations and simulations for sequences 1, 2, 3, and 5 were performed in Matlab.\cite{MATLAB:2010} followed by test and conformation using replicated state-to-state optimal control calculations, while sequence 4 fully optimized using replicated state-to-state optimal control, from a cRW starting guess.\cite{OC_DNP} Calculations that took into account MW inhomogeneity added responses from nine MW field strengths with scaling factors \{0.65, 0.70, 0.75, 0.80, 0.85, 0.90, 0.95, 1.00, 1.05\} and corresponding weights \{0.079, 0.083, 0.088, 0.094, 0.103, 0.115, 0.135, 0.209, 0.095\} in consistency with a power-model to cope with MW field inhomogeneity. \cite{GUPTA201517} 

\section{CONCLUSIONS}
In conclusion, we have addressed fundamental elements in designing broadband dipolar recoupling pulse sequences with focus on pulsed DNP. We have outlined a number of important elements for systematic experiment design including ($i$) careful selection of the resonance condition and pulse sequence modulation frequency,  to avoid destructive overlap of ZQ and DQ mediated oppositely signed polarization transfer, fulfillment of the first-order approximation of the FOM fidelity function, and effective MW fields sufficiently small to reduce effects from experimentally relevant MW field inhomogeneity, ($ii$) a simple, yet very fast constrained random walk (cRW) procedure for direct experiment design or as a tool to provide good starting guesses for subsequent ($iii$) non-linear optimization. The latter is suggested accomplished by designing pulse sequences with sufficiently large effective fields on the involved spin channels and optimizing a figure of merit (FOM) function that balances linear and bilinear terms of the effective Hamiltonian in the basic pulse sequence element to achieve pulse sequences with very large excitation band widths. To illustrate these aspects we have presented a detailed analysis of designed optimal cRW-OPT DNP pulse sequences and compared the theoretical/numerical findings with experimental implementation of cRW-OPT pulse sequences on state-of-the-art pulsed DNP instrumentation. The constrained random walk was implemented for pulse sequences without any phase modulation. This makes it easy to efficiently sample the possible space while ensuring that the sequences are physically attainable in terms of driving field amplitude. In principle, the random walk can also be implemented with phase-modulated pulse sequences. However, in this case, it is far less trivial to efficiently sample the possible sequence space. Solving this problem is the topic of currently ongoing work.

We envisage that the developed DNP pulse sequences can find immediate application for low-field (X-band) DNP taking polarization from unpaired electron spins with broad EPR lines. As demonstrated recently\cite{DNP_SSNMR_Carvalho:2025aa} optimal DNP sequences may easily be translated into heteronuclear MAS dipolar recoupling experiments with unique broadband performance. On a broader horizon the design procedure may find applications for optimizing DNP experiments for applications high-field instrumentation under consideration reduced availability of high-power pulsed MW sources and potentially under influence of magic-angle-spinning. Finally, pulsed DNP has an considerable potential in broader modalities of quantum sensing, where specific pulse sequence engineering as demonstrated in this paper may be of significant interest.

\begin{acknowledgments}

The authors acknowledge advice from Dr. A. Doll (Paul Scherrer Institut, CH), Dr. D. Klose (ETH, Z$\ddot{\text{u}}$rich, CH), and Prof. G. Jeschke (ETH, Z$\ddot{\text{u}}$rich, CH) on building the pulsed X-band EPR/DNP instrumentation. We acknowledge  support from the Aarhus University Research Foundation (AUFF, grant AUFF-E-2021-9-22), the Novo Nordisk Foundation (NERD grant NNF22OC0076002), the Villum Foundation Synergy programme (grant 50099), the Swiss National Science Foundation (Postdoc.Mobility grant 206623),  and the DeiC National HPC (g.a. DeiC-AU-N5-2024094-H2-2024-35).

\end{acknowledgments}

\section*{AUTHOR DECLARATIONS}
	\subsection*{Conflict of Interests:}
		The authors declare that they have no conflicting interests.

  	\subsection*{Author contributions}
All authors contributed to development, discussion, and validation  of the theory. NCN, ABN, NW, JPC, and TU developed the formalism and made the optimization code, supported by FVJ, DLG and ZT. All contributed to the definition of  the scope of the optimizations. Experiments were accomplished by JPC, ABN, and FVJ. All contributed to writing the manuscript.
  
\section*{Data Availability Statement}
The data needed to evaluate the conclusions in the paper are present in the paper. 
Optimization code can be made available upon reasonable request

\makeatletter
\def\bibsection{\relax} 
\makeatother

\section*{References}
\bibliography{RW_lib.bib}

\begin{thebibliography}{60}%
\makeatletter
\providecommand \@ifxundefined [1]{%
 \@ifx{#1\undefined}
}%
\providecommand \@ifnum [1]{%
 \ifnum #1\expandafter \@firstoftwo
 \else \expandafter \@secondoftwo
 \fi
}%
\providecommand \@ifx [1]{%
 \ifx #1\expandafter \@firstoftwo
 \else \expandafter \@secondoftwo
 \fi
}%
\providecommand \natexlab [1]{#1}%
\providecommand \enquote  [1]{``#1''}%
\providecommand \bibnamefont  [1]{#1}%
\providecommand \bibfnamefont [1]{#1}%
\providecommand \citenamefont [1]{#1}%
\providecommand \href@noop [0]{\@secondoftwo}%
\providecommand \href [0]{\begingroup \@sanitize@url \@href}%
\providecommand \@href[1]{\@@startlink{#1}\@@href}%
\providecommand \@@href[1]{\endgroup#1\@@endlink}%
\providecommand \@sanitize@url [0]{\catcode `\\12\catcode `\$12\catcode `\&12\catcode `\#12\catcode `\^12\catcode `\_12\catcode `\%12\relax}%
\providecommand \@@startlink[1]{}%
\providecommand \@@endlink[0]{}%
\providecommand \url  [0]{\begingroup\@sanitize@url \@url }%
\providecommand \@url [1]{\endgroup\@href {#1}{\urlprefix }}%
\providecommand \urlprefix  [0]{URL }%
\providecommand \Eprint [0]{\href }%
\providecommand \doibase [0]{https://doi.org/}%
\providecommand \selectlanguage [0]{\@gobble}%
\providecommand \bibinfo  [0]{\@secondoftwo}%
\providecommand \bibfield  [0]{\@secondoftwo}%
\providecommand \translation [1]{[#1]}%
\providecommand \BibitemOpen [0]{}%
\providecommand \bibitemStop [0]{}%
\providecommand \bibitemNoStop [0]{.\EOS\space}%
\providecommand \EOS [0]{\spacefactor3000\relax}%
\providecommand \BibitemShut  [1]{\csname bibitem#1\endcsname}%
\let\auto@bib@innerbib\@empty
\bibitem [{\citenamefont {Ernst}, \citenamefont {Bodenhausen},\ and\ \citenamefont {Wokaun}(1990)}]{Ernst_book}%
  \BibitemOpen
  \bibfield  {author} {\bibinfo {author} {\bibfnamefont {R.~R.}\ \bibnamefont {Ernst}}, \bibinfo {author} {\bibfnamefont {G.}~\bibnamefont {Bodenhausen}},\ and\ \bibinfo {author} {\bibfnamefont {A.}~\bibnamefont {Wokaun}},\ }\href@noop {} {\emph {\bibinfo {title} {{Principles of Nuclear Magnetic Resonance in One and Two Dimensions}}}}\ (\bibinfo  {publisher} {Oxford University Press},\ \bibinfo {year} {1990})\BibitemShut {NoStop}%
\bibitem [{\citenamefont {Schweiger}\ and\ \citenamefont {Jeschke}(2001)}]{Jeschke_book}%
  \BibitemOpen
  \bibfield  {author} {\bibinfo {author} {\bibfnamefont {A.}~\bibnamefont {Schweiger}}\ and\ \bibinfo {author} {\bibfnamefont {G.}~\bibnamefont {Jeschke}},\ }\href@noop {} {\emph {\bibinfo {title} {Principles of pulse electron paramagnetic resonance}}}\ (\bibinfo  {publisher} {Oxford university press},\ \bibinfo {year} {2001})\BibitemShut {NoStop}%
\bibitem [{\citenamefont {Haeberlen}\ and\ \citenamefont {Waugh}(1968)}]{AHT}%
  \BibitemOpen
  \bibfield  {author} {\bibinfo {author} {\bibfnamefont {U.}~\bibnamefont {Haeberlen}}\ and\ \bibinfo {author} {\bibfnamefont {J.~S.}\ \bibnamefont {Waugh}},\ }\bibfield  {title} {\enquote {\bibinfo {title} {Coherent averaging effects in magnetic resonance},}\ }\href {https://doi.org/10.1103/PhysRev.175.453} {\bibfield  {journal} {\bibinfo  {journal} {Phys. Rev.}\ }\textbf {\bibinfo {volume} {175}},\ \bibinfo {pages} {453--467} (\bibinfo {year} {1968})}\BibitemShut {NoStop}%
\bibitem [{\citenamefont {Maricq}\ and\ \citenamefont {Waugh}(1979)}]{MaricqWaugh}%
  \BibitemOpen
  \bibfield  {author} {\bibinfo {author} {\bibfnamefont {M.}~\bibnamefont {Maricq}}\ and\ \bibinfo {author} {\bibfnamefont {J.~S.}\ \bibnamefont {Waugh}},\ }\bibfield  {title} {\enquote {\bibinfo {title} {Nmr in rotating solids},}\ }\href {https://doi.org/10.1063/1.437915} {\bibfield  {journal} {\bibinfo  {journal} {The Journal of Chemical Physics}\ }\textbf {\bibinfo {volume} {70}},\ \bibinfo {pages} {3300--3316} (\bibinfo {year} {1979})},\ \Eprint {https://arxiv.org/abs/https://doi.org/10.1063/1.437915} {https://doi.org/10.1063/1.437915} \BibitemShut {NoStop}%
\bibitem [{\citenamefont {Hohwy}\ and\ \citenamefont {Nielsen}(1998)}]{scBCH}%
  \BibitemOpen
  \bibfield  {author} {\bibinfo {author} {\bibfnamefont {M.}~\bibnamefont {Hohwy}}\ and\ \bibinfo {author} {\bibfnamefont {N.~C.}\ \bibnamefont {Nielsen}},\ }\bibfield  {title} {\enquote {\bibinfo {title} {Systematic design and evaluation of multiple-pulse experiments in nuclear magnetic resonance spectroscopy using a semi-continuous baker--campbell--hausdorff expansion},}\ }\href {https://doi.org/10.1063/1.476978} {\bibfield  {journal} {\bibinfo  {journal} {The Journal of Chemical Physics}\ }\textbf {\bibinfo {volume} {109}},\ \bibinfo {pages} {3780--3791} (\bibinfo {year} {1998})},\ \Eprint {https://arxiv.org/abs/https://doi.org/10.1063/1.476978} {https://doi.org/10.1063/1.476978} \BibitemShut {NoStop}%
\bibitem [{\citenamefont {S{\o}rensen}\ \emph {et~al.}(1984)\citenamefont {S{\o}rensen}, \citenamefont {Eich}, \citenamefont {Levitt}, \citenamefont {Bodenhausen},\ and\ \citenamefont {Ernst}}]{Prodop}%
  \BibitemOpen
  \bibfield  {author} {\bibinfo {author} {\bibfnamefont {O.}~\bibnamefont {S{\o}rensen}}, \bibinfo {author} {\bibfnamefont {G.}~\bibnamefont {Eich}}, \bibinfo {author} {\bibfnamefont {M.}~\bibnamefont {Levitt}}, \bibinfo {author} {\bibfnamefont {G.}~\bibnamefont {Bodenhausen}},\ and\ \bibinfo {author} {\bibfnamefont {R.}~\bibnamefont {Ernst}},\ }\bibfield  {title} {\enquote {\bibinfo {title} {Product operator formalism for the description of nmr pulse experiments},}\ }\href {https://doi.org/https://doi.org/10.1016/0079-6565(84)80005-9} {\bibfield  {journal} {\bibinfo  {journal} {Progress in Nuclear Magnetic Resonance Spectroscopy}\ }\textbf {\bibinfo {volume} {16}},\ \bibinfo {pages} {163--192} (\bibinfo {year} {1984})}\BibitemShut {NoStop}%
\bibitem [{\citenamefont {Shirley}(1965)}]{Shirley}%
  \BibitemOpen
  \bibfield  {author} {\bibinfo {author} {\bibfnamefont {J.~H.}\ \bibnamefont {Shirley}},\ }\bibfield  {title} {\enquote {\bibinfo {title} {Solution of the schr\"odinger equation with a hamiltonian periodic in time},}\ }\href {https://doi.org/10.1103/PhysRev.138.B979} {\bibfield  {journal} {\bibinfo  {journal} {Phys. Rev.}\ }\textbf {\bibinfo {volume} {138}},\ \bibinfo {pages} {B979--B987} (\bibinfo {year} {1965})}\BibitemShut {NoStop}%
\bibitem [{\citenamefont {Weintraub}\ and\ \citenamefont {Vega}(1993)}]{Weintraub}%
  \BibitemOpen
  \bibfield  {author} {\bibinfo {author} {\bibfnamefont {O.}~\bibnamefont {Weintraub}}\ and\ \bibinfo {author} {\bibfnamefont {S.}~\bibnamefont {Vega}},\ }\bibfield  {title} {\enquote {\bibinfo {title} {Floquet density matrices and effective hamiltonians in magic-angle-spinning nmr spectroscopy},}\ }\href {https://doi.org/https://doi.org/10.1006/jmra.1993.1279} {\bibfield  {journal} {\bibinfo  {journal} {Journal of Magnetic Resonance, Series A}\ }\textbf {\bibinfo {volume} {105}},\ \bibinfo {pages} {245--267} (\bibinfo {year} {1993})}\BibitemShut {NoStop}%
\bibitem [{\citenamefont {Scholz}, \citenamefont {{van Beek}},\ and\ \citenamefont {Ernst}(2010)}]{Scholz2010}%
  \BibitemOpen
  \bibfield  {author} {\bibinfo {author} {\bibfnamefont {I.}~\bibnamefont {Scholz}}, \bibinfo {author} {\bibfnamefont {J.~D.}\ \bibnamefont {{van Beek}}},\ and\ \bibinfo {author} {\bibfnamefont {M.}~\bibnamefont {Ernst}},\ }\bibfield  {title} {\enquote {\bibinfo {title} {Operator-based floquet theory in solid-state nmr},}\ }\href {https://doi.org/https://doi.org/10.1016/j.ssnmr.2010.04.003} {\bibfield  {journal} {\bibinfo  {journal} {Solid State Nuclear Magnetic Resonance}\ }\textbf {\bibinfo {volume} {37}},\ \bibinfo {pages} {39--59} (\bibinfo {year} {2010})}\BibitemShut {NoStop}%
\bibitem [{\citenamefont {Untidt}\ and\ \citenamefont {Nielsen}(2002)}]{EEHT}%
  \BibitemOpen
  \bibfield  {author} {\bibinfo {author} {\bibfnamefont {T.~S.}\ \bibnamefont {Untidt}}\ and\ \bibinfo {author} {\bibfnamefont {N.~C.}\ \bibnamefont {Nielsen}},\ }\bibfield  {title} {\enquote {\bibinfo {title} {Closed solution to the baker-campbell-hausdorff problem: Exact effective hamiltonian theory for analysis of nuclear-magnetic-resonance experiments},}\ }\href {https://doi.org/10.1103/PhysRevE.65.021108} {\bibfield  {journal} {\bibinfo  {journal} {Phys. Rev. E}\ }\textbf {\bibinfo {volume} {65}},\ \bibinfo {pages} {021108} (\bibinfo {year} {2002})}\BibitemShut {NoStop}%
\bibitem [{\citenamefont {Siminovitch}, \citenamefont {Untidt},\ and\ \citenamefont {Nielsen}(2004)}]{EEHT2}%
  \BibitemOpen
  \bibfield  {author} {\bibinfo {author} {\bibfnamefont {D.}~\bibnamefont {Siminovitch}}, \bibinfo {author} {\bibfnamefont {T.}~\bibnamefont {Untidt}},\ and\ \bibinfo {author} {\bibfnamefont {N.~C.}\ \bibnamefont {Nielsen}},\ }\bibfield  {title} {\enquote {\bibinfo {title} {Exact effective hamiltonian theory. ii. polynomial expansion of matrix functions and entangled unitary exponential operators},}\ }\href {https://doi.org/10.1063/1.1628216} {\bibfield  {journal} {\bibinfo  {journal} {The Journal of Chemical Physics}\ }\textbf {\bibinfo {volume} {120}},\ \bibinfo {pages} {51--66} (\bibinfo {year} {2004})},\ \Eprint {https://arxiv.org/abs/https://doi.org/10.1063/1.1628216} {https://doi.org/10.1063/1.1628216} \BibitemShut {NoStop}%
\bibitem [{\citenamefont {Shankar}\ \emph {et~al.}(2017)\citenamefont {Shankar}, \citenamefont {Ernst}, \citenamefont {Madhu}, \citenamefont {Vosegaard}, \citenamefont {Nielsen},\ and\ \citenamefont {Nielsen}}]{shankar}%
  \BibitemOpen
  \bibfield  {author} {\bibinfo {author} {\bibfnamefont {R.}~\bibnamefont {Shankar}}, \bibinfo {author} {\bibfnamefont {M.}~\bibnamefont {Ernst}}, \bibinfo {author} {\bibfnamefont {P.~K.}\ \bibnamefont {Madhu}}, \bibinfo {author} {\bibfnamefont {T.}~\bibnamefont {Vosegaard}}, \bibinfo {author} {\bibfnamefont {N.~C.}\ \bibnamefont {Nielsen}},\ and\ \bibinfo {author} {\bibfnamefont {A.~B.}\ \bibnamefont {Nielsen}},\ }\bibfield  {title} {\enquote {\bibinfo {title} {A general theoretical description of the influence of isotropic chemical shift in dipolar recoupling experiments for solid-state nmr},}\ }\href {https://doi.org/10.1063/1.4979123} {\bibfield  {journal} {\bibinfo  {journal} {The Journal of Chemical Physics}\ }\textbf {\bibinfo {volume} {146}},\ \bibinfo {pages} {134105} (\bibinfo {year} {2017})},\ \Eprint {https://arxiv.org/abs/https://doi.org/10.1063/1.4979123} {https://doi.org/10.1063/1.4979123} \BibitemShut {NoStop}%
\bibitem [{\citenamefont {Nielsen}\ \emph {et~al.}(2019)\citenamefont {Nielsen}, \citenamefont {Hansen}, \citenamefont {Andersen},\ and\ \citenamefont {Vosegaard}}]{SSV-EHT}%
  \BibitemOpen
  \bibfield  {author} {\bibinfo {author} {\bibfnamefont {A.~B.}\ \bibnamefont {Nielsen}}, \bibinfo {author} {\bibfnamefont {M.~R.}\ \bibnamefont {Hansen}}, \bibinfo {author} {\bibfnamefont {J.~E.}\ \bibnamefont {Andersen}},\ and\ \bibinfo {author} {\bibfnamefont {T.}~\bibnamefont {Vosegaard}},\ }\bibfield  {title} {\enquote {\bibinfo {title} {Single-spin vector analysis of strongly coupled nuclei in tocsy nmr experiments},}\ }\href {https://doi.org/10.1063/1.5123046} {\bibfield  {journal} {\bibinfo  {journal} {The Journal of Chemical Physics}\ }\textbf {\bibinfo {volume} {151}},\ \bibinfo {pages} {134117} (\bibinfo {year} {2019})},\ \Eprint {https://arxiv.org/abs/https://doi.org/10.1063/1.5123046} {https://doi.org/10.1063/1.5123046} \BibitemShut {NoStop}%
\bibitem [{\citenamefont {Nielsen}\ and\ \citenamefont {Nielsen}(2022)}]{SVEHT_EEHT}%
  \BibitemOpen
  \bibfield  {author} {\bibinfo {author} {\bibfnamefont {A.~B.}\ \bibnamefont {Nielsen}}\ and\ \bibinfo {author} {\bibfnamefont {N.~C.}\ \bibnamefont {Nielsen}},\ }\bibfield  {title} {\enquote {\bibinfo {title} {Accurate analysis and perspectives for systematic design of magnetic resonance experiments using single-spin vector and exact effective hamiltonian theory},}\ }\href {https://doi.org/https://doi.org/10.1016/j.jmro.2022.100064} {\bibfield  {journal} {\bibinfo  {journal} {Journal of Magnetic Resonance Open}\ }\textbf {\bibinfo {volume} {12-13}},\ \bibinfo {pages} {100064} (\bibinfo {year} {2022})}\BibitemShut {NoStop}%
\bibitem [{\citenamefont {Bak}, \citenamefont {Rasmussen},\ and\ \citenamefont {Nielsen}(2000)}]{SIMPSON}%
  \BibitemOpen
  \bibfield  {author} {\bibinfo {author} {\bibfnamefont {M.}~\bibnamefont {Bak}}, \bibinfo {author} {\bibfnamefont {J.~T.}\ \bibnamefont {Rasmussen}},\ and\ \bibinfo {author} {\bibfnamefont {N.~C.}\ \bibnamefont {Nielsen}},\ }\bibfield  {title} {\enquote {\bibinfo {title} {Simpson: A general simulation program for solid-state nmr spectroscopy},}\ }\href {https://doi.org/https://doi.org/10.1006/jmre.2000.2179} {\bibfield  {journal} {\bibinfo  {journal} {Journal of Magnetic Resonance}\ }\textbf {\bibinfo {volume} {147}},\ \bibinfo {pages} {296--330} (\bibinfo {year} {2000})}\BibitemShut {NoStop}%
\bibitem [{\citenamefont {Veshtort}\ and\ \citenamefont {Griffin}(2006)}]{spinevolution}%
  \BibitemOpen
  \bibfield  {author} {\bibinfo {author} {\bibfnamefont {M.}~\bibnamefont {Veshtort}}\ and\ \bibinfo {author} {\bibfnamefont {R.~G.}\ \bibnamefont {Griffin}},\ }\bibfield  {title} {\enquote {\bibinfo {title} {Spinevolution: A powerful tool for the simulation of solid and liquid state nmr experiments},}\ }\href {https://doi.org/https://doi.org/10.1016/j.jmr.2005.07.018} {\bibfield  {journal} {\bibinfo  {journal} {Journal of Magnetic Resonance}\ }\textbf {\bibinfo {volume} {178}},\ \bibinfo {pages} {248--282} (\bibinfo {year} {2006})}\BibitemShut {NoStop}%
\bibitem [{\citenamefont {Hogben}\ \emph {et~al.}(2011)\citenamefont {Hogben}, \citenamefont {Krzystyniak}, \citenamefont {Charnock}, \citenamefont {Hore},\ and\ \citenamefont {Kuprov}}]{spinach}%
  \BibitemOpen
  \bibfield  {author} {\bibinfo {author} {\bibfnamefont {H.}~\bibnamefont {Hogben}}, \bibinfo {author} {\bibfnamefont {M.}~\bibnamefont {Krzystyniak}}, \bibinfo {author} {\bibfnamefont {G.}~\bibnamefont {Charnock}}, \bibinfo {author} {\bibfnamefont {P.}~\bibnamefont {Hore}},\ and\ \bibinfo {author} {\bibfnamefont {I.}~\bibnamefont {Kuprov}},\ }\bibfield  {title} {\enquote {\bibinfo {title} {Spinach -- a software library for simulation of spin dynamics in large spin systems},}\ }\href {https://doi.org/https://doi.org/10.1016/j.jmr.2010.11.008} {\bibfield  {journal} {\bibinfo  {journal} {Journal of Magnetic Resonance}\ }\textbf {\bibinfo {volume} {208}},\ \bibinfo {pages} {179--194} (\bibinfo {year} {2011})}\BibitemShut {NoStop}%
\bibitem [{\citenamefont {Press}\ \emph {et~al.}(2007)\citenamefont {Press}, \citenamefont {Teukolsky}, \citenamefont {Vetterling},\ and\ \citenamefont {Flannery}}]{press2007numerical}%
  \BibitemOpen
  \bibfield  {author} {\bibinfo {author} {\bibfnamefont {W.}~\bibnamefont {Press}}, \bibinfo {author} {\bibfnamefont {S.}~\bibnamefont {Teukolsky}}, \bibinfo {author} {\bibfnamefont {W.}~\bibnamefont {Vetterling}},\ and\ \bibinfo {author} {\bibfnamefont {B.}~\bibnamefont {Flannery}},\ }\href@noop {} {\emph {\bibinfo {title} {Numerical Recipes: The Art of Scientific Computing}}},\ \bibinfo {edition} {3rd}\ ed.\ (\bibinfo  {publisher} {Cambridge University Press},\ \bibinfo {year} {2007})\BibitemShut {NoStop}%
\bibitem [{\citenamefont {Pontryagin}\ \emph {et~al.}(1962)\citenamefont {Pontryagin}, \citenamefont {Boltayanskii}, \citenamefont {Gamkrelidze},\ and\ \citenamefont {Mishchenko}}]{Pontryagin}%
  \BibitemOpen
  \bibfield  {author} {\bibinfo {author} {\bibfnamefont {L.}~\bibnamefont {Pontryagin}}, \bibinfo {author} {\bibfnamefont {V.}~\bibnamefont {Boltayanskii}}, \bibinfo {author} {\bibfnamefont {R.}~\bibnamefont {Gamkrelidze}},\ and\ \bibinfo {author} {\bibfnamefont {E.}~\bibnamefont {Mishchenko}},\ }\href@noop {} {\emph {\bibinfo {title} {The mathematical theory of optimal processes}}}\ (\bibinfo  {publisher} {Wiley},\ \bibinfo {year} {1962})\BibitemShut {NoStop}%
\bibitem [{\citenamefont {Krotov}(1996)}]{Krotov}%
  \BibitemOpen
  \bibfield  {author} {\bibinfo {author} {\bibfnamefont {V.~F.}\ \bibnamefont {Krotov}},\ }\href@noop {} {\emph {\bibinfo {title} {Global methods in optimal control theory}}}\ (\bibinfo  {publisher} {Marcel Dekker, New York},\ \bibinfo {year} {1996})\BibitemShut {NoStop}%
\bibitem [{\citenamefont {Morzhin}\ and\ \citenamefont {Pechen}(2019)}]{Morzhin_2019}%
  \BibitemOpen
  \bibfield  {author} {\bibinfo {author} {\bibfnamefont {O.~V.}\ \bibnamefont {Morzhin}}\ and\ \bibinfo {author} {\bibfnamefont {A.~N.}\ \bibnamefont {Pechen}},\ }\bibfield  {title} {\enquote {\bibinfo {title} {Krotov method for optimal control of closed quantum systems},}\ }\href {https://doi.org/10.1070/RM9835} {\bibfield  {journal} {\bibinfo  {journal} {Russian Mathematical Surveys}\ }\textbf {\bibinfo {volume} {74}},\ \bibinfo {pages} {851} (\bibinfo {year} {2019})}\BibitemShut {NoStop}%
\bibitem [{\citenamefont {Khaneja}, \citenamefont {Brockett},\ and\ \citenamefont {Glaser}(2001)}]{OC1}%
  \BibitemOpen
  \bibfield  {author} {\bibinfo {author} {\bibfnamefont {N.}~\bibnamefont {Khaneja}}, \bibinfo {author} {\bibfnamefont {R.}~\bibnamefont {Brockett}},\ and\ \bibinfo {author} {\bibfnamefont {S.~J.}\ \bibnamefont {Glaser}},\ }\bibfield  {title} {\enquote {\bibinfo {title} {Time optimal control in spin systems},}\ }\href {https://doi.org/10.1103/PhysRevA.63.032308} {\bibfield  {journal} {\bibinfo  {journal} {Phys. Rev. A}\ }\textbf {\bibinfo {volume} {63}},\ \bibinfo {pages} {032308} (\bibinfo {year} {2001})}\BibitemShut {NoStop}%
\bibitem [{\citenamefont {Khaneja}\ \emph {et~al.}(2005)\citenamefont {Khaneja}, \citenamefont {Reiss}, \citenamefont {Kehlet}, \citenamefont {Schulte-Herbr{\"u}ggen},\ and\ \citenamefont {Glaser}}]{grape}%
  \BibitemOpen
  \bibfield  {author} {\bibinfo {author} {\bibfnamefont {N.}~\bibnamefont {Khaneja}}, \bibinfo {author} {\bibfnamefont {T.}~\bibnamefont {Reiss}}, \bibinfo {author} {\bibfnamefont {C.}~\bibnamefont {Kehlet}}, \bibinfo {author} {\bibfnamefont {T.}~\bibnamefont {Schulte-Herbr{\"u}ggen}},\ and\ \bibinfo {author} {\bibfnamefont {S.~J.}\ \bibnamefont {Glaser}},\ }\bibfield  {title} {\enquote {\bibinfo {title} {Optimal control of coupled spin dynamics: design of nmr pulse sequences by gradient ascent algorithms},}\ }\href {https://doi.org/https://doi.org/10.1016/j.jmr.2004.11.004} {\bibfield  {journal} {\bibinfo  {journal} {Journal of Magnetic Resonance}\ }\textbf {\bibinfo {volume} {172}},\ \bibinfo {pages} {296--305} (\bibinfo {year} {2005})}\BibitemShut {NoStop}%
\bibitem [{\citenamefont {Kehlet}\ \emph {et~al.}(2004)\citenamefont {Kehlet}, \citenamefont {Sivertsen}, \citenamefont {Bjerring}, \citenamefont {Reiss}, \citenamefont {Khaneja}, \citenamefont {Glaser},\ and\ \citenamefont {Nielsen}}]{OCsolid}%
  \BibitemOpen
  \bibfield  {author} {\bibinfo {author} {\bibfnamefont {C.~T.}\ \bibnamefont {Kehlet}}, \bibinfo {author} {\bibfnamefont {A.~C.}\ \bibnamefont {Sivertsen}}, \bibinfo {author} {\bibfnamefont {M.}~\bibnamefont {Bjerring}}, \bibinfo {author} {\bibfnamefont {T.~O.}\ \bibnamefont {Reiss}}, \bibinfo {author} {\bibfnamefont {N.}~\bibnamefont {Khaneja}}, \bibinfo {author} {\bibfnamefont {S.~J.}\ \bibnamefont {Glaser}},\ and\ \bibinfo {author} {\bibfnamefont {N.~C.}\ \bibnamefont {Nielsen}},\ }\bibfield  {title} {\enquote {\bibinfo {title} {Improving solid-state nmr dipolar recoupling by optimal control},}\ }\href {https://doi.org/10.1021/ja048786e} {\bibfield  {journal} {\bibinfo  {journal} {Journal of the American Chemical Society}\ }\textbf {\bibinfo {volume} {126}},\ \bibinfo {pages} {10202--10203} (\bibinfo {year} {2004})},\ \bibinfo {note} {pMID: 15315406},\ \Eprint {https://arxiv.org/abs/https://doi.org/10.1021/ja048786e} {https://doi.org/10.1021/ja048786e} \BibitemShut {NoStop}%
\bibitem [{\citenamefont {Maximov}, \citenamefont {To{\v s}ner},\ and\ \citenamefont {Nielsen}(2008)}]{MaximovOC}%
  \BibitemOpen
  \bibfield  {author} {\bibinfo {author} {\bibfnamefont {I.~I.}\ \bibnamefont {Maximov}}, \bibinfo {author} {\bibfnamefont {Z.}~\bibnamefont {To{\v s}ner}},\ and\ \bibinfo {author} {\bibfnamefont {N.~C.}\ \bibnamefont {Nielsen}},\ }\bibfield  {title} {\enquote {\bibinfo {title} {Optimal control design of nmr and dynamic nuclear polarization experiments using monotonically convergent algorithms},}\ }\href {https://doi.org/10.1063/1.2903458} {\bibfield  {journal} {\bibinfo  {journal} {The Journal of Chemical Physics}\ }\textbf {\bibinfo {volume} {128}},\ \bibinfo {pages} {184505} (\bibinfo {year} {2008})},\ \Eprint {https://arxiv.org/abs/https://doi.org/10.1063/1.2903458} {https://doi.org/10.1063/1.2903458} \BibitemShut {NoStop}%
\bibitem [{\citenamefont {Kehlet}\ \emph {et~al.}(2007)\citenamefont {Kehlet}, \citenamefont {Bjerring}, \citenamefont {Sivertsen}, \citenamefont {Kristensen}, \citenamefont {Enghild}, \citenamefont {Glaser}, \citenamefont {Khaneja},\ and\ \citenamefont {Nielsen}}]{OCNCONCA}%
  \BibitemOpen
  \bibfield  {author} {\bibinfo {author} {\bibfnamefont {C.}~\bibnamefont {Kehlet}}, \bibinfo {author} {\bibfnamefont {M.}~\bibnamefont {Bjerring}}, \bibinfo {author} {\bibfnamefont {A.~C.}\ \bibnamefont {Sivertsen}}, \bibinfo {author} {\bibfnamefont {T.}~\bibnamefont {Kristensen}}, \bibinfo {author} {\bibfnamefont {J.~J.}\ \bibnamefont {Enghild}}, \bibinfo {author} {\bibfnamefont {S.~J.}\ \bibnamefont {Glaser}}, \bibinfo {author} {\bibfnamefont {N.}~\bibnamefont {Khaneja}},\ and\ \bibinfo {author} {\bibfnamefont {N.~C.}\ \bibnamefont {Nielsen}},\ }\bibfield  {title} {\enquote {\bibinfo {title} {Optimal control based nco and nca experiments for spectral assignment in biological solid-state nmr spectroscopy},}\ }\href {https://doi.org/https://doi.org/10.1016/j.jmr.2007.06.011} {\bibfield  {journal} {\bibinfo  {journal} {Journal of Magnetic Resonance}\ }\textbf {\bibinfo {volume} {188}},\ \bibinfo {pages} {216--230} (\bibinfo {year} {2007})}\BibitemShut {NoStop}%
\bibitem [{\citenamefont {To{\v s}ner}\ \emph {et~al.}(2009)\citenamefont {To{\v s}ner}, \citenamefont {Vosegaard}, \citenamefont {Kehlet}, \citenamefont {Khaneja}, \citenamefont {Glaser},\ and\ \citenamefont {Nielsen}}]{SIMPSONOC}%
  \BibitemOpen
  \bibfield  {author} {\bibinfo {author} {\bibfnamefont {Z.}~\bibnamefont {To{\v s}ner}}, \bibinfo {author} {\bibfnamefont {T.}~\bibnamefont {Vosegaard}}, \bibinfo {author} {\bibfnamefont {C.}~\bibnamefont {Kehlet}}, \bibinfo {author} {\bibfnamefont {N.}~\bibnamefont {Khaneja}}, \bibinfo {author} {\bibfnamefont {S.~J.}\ \bibnamefont {Glaser}},\ and\ \bibinfo {author} {\bibfnamefont {N.~C.}\ \bibnamefont {Nielsen}},\ }\bibfield  {title} {\enquote {\bibinfo {title} {Optimal control in nmr spectroscopy: Numerical implementation in simpson},}\ }\href {https://doi.org/https://doi.org/10.1016/j.jmr.2008.11.020} {\bibfield  {journal} {\bibinfo  {journal} {Journal of Magnetic Resonance}\ }\textbf {\bibinfo {volume} {197}},\ \bibinfo {pages} {120--134} (\bibinfo {year} {2009})}\BibitemShut {NoStop}%
\bibitem [{\citenamefont {Maximov}\ \emph {et~al.}(2010)\citenamefont {Maximov}, \citenamefont {Salomon}, \citenamefont {Turinici},\ and\ \citenamefont {Nielsen}}]{MaximovSmoothing}%
  \BibitemOpen
  \bibfield  {author} {\bibinfo {author} {\bibfnamefont {I.~I.}\ \bibnamefont {Maximov}}, \bibinfo {author} {\bibfnamefont {J.}~\bibnamefont {Salomon}}, \bibinfo {author} {\bibfnamefont {G.}~\bibnamefont {Turinici}},\ and\ \bibinfo {author} {\bibfnamefont {N.~C.}\ \bibnamefont {Nielsen}},\ }\bibfield  {title} {\enquote {\bibinfo {title} {A smoothing monotonic convergent optimal control algorithm for nuclear magnetic resonance pulse sequence design},}\ }\href {https://doi.org/10.1063/1.3328783} {\bibfield  {journal} {\bibinfo  {journal} {The Journal of Chemical Physics}\ }\textbf {\bibinfo {volume} {132}},\ \bibinfo {pages} {084107} (\bibinfo {year} {2010})},\ \Eprint {https://arxiv.org/abs/https://doi.org/10.1063/1.3328783} {https://doi.org/10.1063/1.3328783} \BibitemShut {NoStop}%
\bibitem [{\citenamefont {Vinding}\ \emph {et~al.}(2012)\citenamefont {Vinding}, \citenamefont {Maximov}, \citenamefont {To{\v s}ner},\ and\ \citenamefont {Nielsen}}]{OCMRI_Krotov}%
  \BibitemOpen
  \bibfield  {author} {\bibinfo {author} {\bibfnamefont {M.~S.}\ \bibnamefont {Vinding}}, \bibinfo {author} {\bibfnamefont {I.~I.}\ \bibnamefont {Maximov}}, \bibinfo {author} {\bibfnamefont {Z.}~\bibnamefont {To{\v s}ner}},\ and\ \bibinfo {author} {\bibfnamefont {N.~C.}\ \bibnamefont {Nielsen}},\ }\bibfield  {title} {\enquote {\bibinfo {title} {Fast numerical design of spatial-selective rf pulses in mri using krotov and quasi-newton based optimal control methods},}\ }\href {https://doi.org/10.1063/1.4739755} {\bibfield  {journal} {\bibinfo  {journal} {The Journal of Chemical Physics}\ }\textbf {\bibinfo {volume} {137}},\ \bibinfo {pages} {054203} (\bibinfo {year} {2012})},\ \Eprint {https://arxiv.org/abs/https://doi.org/10.1063/1.4739755} {https://doi.org/10.1063/1.4739755} \BibitemShut {NoStop}%
\bibitem [{\citenamefont {Goodwin}\ and\ \citenamefont {Kuprov}(2016)}]{OCKuprov}%
  \BibitemOpen
  \bibfield  {author} {\bibinfo {author} {\bibfnamefont {D.~L.}\ \bibnamefont {Goodwin}}\ and\ \bibinfo {author} {\bibfnamefont {I.}~\bibnamefont {Kuprov}},\ }\bibfield  {title} {\enquote {\bibinfo {title} {Modified newton-raphson grape methods for optimal control of spin systems},}\ }\href {https://doi.org/10.1063/1.4949534} {\bibfield  {journal} {\bibinfo  {journal} {The Journal of Chemical Physics}\ }\textbf {\bibinfo {volume} {144}},\ \bibinfo {pages} {204107} (\bibinfo {year} {2016})},\ \Eprint {https://arxiv.org/abs/https://doi.org/10.1063/1.4949534} {https://doi.org/10.1063/1.4949534} \BibitemShut {NoStop}%
\bibitem [{\citenamefont {To{\v s}ner}\ \emph {et~al.}(2018)\citenamefont {To{\v s}ner}, \citenamefont {Sarkar}, \citenamefont {Becker-Baldus}, \citenamefont {Glaubitz}, \citenamefont {Wegner}, \citenamefont {Engelke}, \citenamefont {Glaser},\ and\ \citenamefont {Reif}}]{Tosner:2018tb}%
  \BibitemOpen
  \bibfield  {author} {\bibinfo {author} {\bibfnamefont {Z.}~\bibnamefont {To{\v s}ner}}, \bibinfo {author} {\bibfnamefont {R.}~\bibnamefont {Sarkar}}, \bibinfo {author} {\bibfnamefont {J.}~\bibnamefont {Becker-Baldus}}, \bibinfo {author} {\bibfnamefont {C.}~\bibnamefont {Glaubitz}}, \bibinfo {author} {\bibfnamefont {S.}~\bibnamefont {Wegner}}, \bibinfo {author} {\bibfnamefont {F.}~\bibnamefont {Engelke}}, \bibinfo {author} {\bibfnamefont {S.~J.}\ \bibnamefont {Glaser}},\ and\ \bibinfo {author} {\bibfnamefont {B.}~\bibnamefont {Reif}},\ }\bibfield  {title} {\enquote {\bibinfo {title} {Overcoming volume selectivity of dipolar recoupling in biological solid-state nmr spectroscopy.}}\ }\href {https://doi.org/10.1002/anie.201805002} {\bibfield  {journal} {\bibinfo  {journal} {Angew Chem Int Ed Engl}\ }\textbf {\bibinfo {volume} {57}},\ \bibinfo {pages} {14514--14518} (\bibinfo {year} {2018})}\BibitemShut {NoStop}%
\bibitem [{\citenamefont {To{\v s}ner}\ \emph {et~al.}(2021)\citenamefont {To{\v s}ner}, \citenamefont {Brandl}, \citenamefont {Blahut}, \citenamefont {Glaser},\ and\ \citenamefont {Reif}}]{TosnerSciAdv}%
  \BibitemOpen
  \bibfield  {author} {\bibinfo {author} {\bibfnamefont {Z.}~\bibnamefont {To{\v s}ner}}, \bibinfo {author} {\bibfnamefont {M.~J.}\ \bibnamefont {Brandl}}, \bibinfo {author} {\bibfnamefont {J.}~\bibnamefont {Blahut}}, \bibinfo {author} {\bibfnamefont {S.~J.}\ \bibnamefont {Glaser}},\ and\ \bibinfo {author} {\bibfnamefont {B.}~\bibnamefont {Reif}},\ }\bibfield  {title} {\enquote {\bibinfo {title} {Maximizing efficiency of dipolar recoupling in solid-state nmr using optimal control sequences},}\ }\href {https://doi.org/10.1126/sciadv.abj5913} {\bibfield  {journal} {\bibinfo  {journal} {Science Advances}\ }\textbf {\bibinfo {volume} {7}},\ \bibinfo {pages} {eabj5913} (\bibinfo {year} {2021})},\ \Eprint {https://arxiv.org/abs/https://www.science.org/doi/pdf/10.1126/sciadv.abj5913} {https://www.science.org/doi/pdf/10.1126/sciadv.abj5913} \BibitemShut {NoStop}%
\bibitem [{\citenamefont {Henstra}\ \emph {et~al.}(1988)\citenamefont {Henstra}, \citenamefont {Dirksen}, \citenamefont {Schmidt},\ and\ \citenamefont {Wenckebach}}]{NOVEL}%
  \BibitemOpen
  \bibfield  {author} {\bibinfo {author} {\bibfnamefont {A.}~\bibnamefont {Henstra}}, \bibinfo {author} {\bibfnamefont {P.}~\bibnamefont {Dirksen}}, \bibinfo {author} {\bibfnamefont {J.}~\bibnamefont {Schmidt}},\ and\ \bibinfo {author} {\bibfnamefont {W.}~\bibnamefont {Wenckebach}},\ }\bibfield  {title} {\enquote {\bibinfo {title} {Nuclear spin orientation via electron spin locking (novel)},}\ }\href {https://doi.org/https://doi.org/10.1016/0022-2364(88)90190-4} {\bibfield  {journal} {\bibinfo  {journal} {Journal of Magnetic Resonance (1969)}\ }\textbf {\bibinfo {volume} {77}},\ \bibinfo {pages} {389--393} (\bibinfo {year} {1988})}\BibitemShut {NoStop}%
\bibitem [{\citenamefont {Tan}\ \emph {et~al.}(2019)\citenamefont {Tan}, \citenamefont {Yang}, \citenamefont {Weber}, \citenamefont {Mathies},\ and\ \citenamefont {Griffin}}]{TOP_DNP}%
  \BibitemOpen
  \bibfield  {author} {\bibinfo {author} {\bibfnamefont {K.~O.}\ \bibnamefont {Tan}}, \bibinfo {author} {\bibfnamefont {C.}~\bibnamefont {Yang}}, \bibinfo {author} {\bibfnamefont {R.~T.}\ \bibnamefont {Weber}}, \bibinfo {author} {\bibfnamefont {G.}~\bibnamefont {Mathies}},\ and\ \bibinfo {author} {\bibfnamefont {R.~G.}\ \bibnamefont {Griffin}},\ }\bibfield  {title} {\enquote {\bibinfo {title} {Time-optimized pulsed dynamic nuclear polarization},}\ }\href@noop {} {\bibfield  {journal} {\bibinfo  {journal} {Science Advances}\ }\textbf {\bibinfo {volume} {5}},\ \bibinfo {pages} {eaav6909} (\bibinfo {year} {2019})}\BibitemShut {NoStop}%
\bibitem [{\citenamefont {Wili}\ \emph {et~al.}(2022)\citenamefont {Wili}, \citenamefont {Nielsen}, \citenamefont {V{\"o}lker}, \citenamefont {Schreder}, \citenamefont {Nielsen}, \citenamefont {Jeschke},\ and\ \citenamefont {Tan}}]{BEAM}%
  \BibitemOpen
  \bibfield  {author} {\bibinfo {author} {\bibfnamefont {N.}~\bibnamefont {Wili}}, \bibinfo {author} {\bibfnamefont {A.~B.}\ \bibnamefont {Nielsen}}, \bibinfo {author} {\bibfnamefont {L.~A.}\ \bibnamefont {V{\"o}lker}}, \bibinfo {author} {\bibfnamefont {L.}~\bibnamefont {Schreder}}, \bibinfo {author} {\bibfnamefont {N.~C.}\ \bibnamefont {Nielsen}}, \bibinfo {author} {\bibfnamefont {G.}~\bibnamefont {Jeschke}},\ and\ \bibinfo {author} {\bibfnamefont {K.~O.}\ \bibnamefont {Tan}},\ }\bibfield  {title} {\enquote {\bibinfo {title} {Designing broadband pulsed dynamic nuclear polarization sequences in static solids},}\ }\href@noop {} {\bibfield  {journal} {\bibinfo  {journal} {Science Advances}\ }\textbf {\bibinfo {volume} {8}},\ \bibinfo {pages} {eabq0536} (\bibinfo {year} {2022})}\BibitemShut {NoStop}%
\bibitem [{\citenamefont {Redrouthu}\ and\ \citenamefont {Mathies}(2022)}]{XiX_DNP}%
  \BibitemOpen
  \bibfield  {author} {\bibinfo {author} {\bibfnamefont {V.~S.}\ \bibnamefont {Redrouthu}}\ and\ \bibinfo {author} {\bibfnamefont {G.}~\bibnamefont {Mathies}},\ }\bibfield  {title} {\enquote {\bibinfo {title} {Efficient pulsed dynamic nuclear polarization with the x-inverse-x sequence},}\ }\href@noop {} {\bibfield  {journal} {\bibinfo  {journal} {Journal of the American Chemical Society}\ }\textbf {\bibinfo {volume} {144}},\ \bibinfo {pages} {1513--1516} (\bibinfo {year} {2022})}\BibitemShut {NoStop}%
\bibitem [{\citenamefont {Redrouthu}, \citenamefont {Vinod-Kumar},\ and\ \citenamefont {Mathies}(2023)}]{TPPM_DNP}%
  \BibitemOpen
  \bibfield  {author} {\bibinfo {author} {\bibfnamefont {V.~S.}\ \bibnamefont {Redrouthu}}, \bibinfo {author} {\bibfnamefont {S.}~\bibnamefont {Vinod-Kumar}},\ and\ \bibinfo {author} {\bibfnamefont {G.}~\bibnamefont {Mathies}},\ }\bibfield  {title} {\enquote {\bibinfo {title} {{Dynamic nuclear polarization by two-pulse phase modulation}},}\ }\href@noop {} {\bibfield  {journal} {\bibinfo  {journal} {The Journal of Chemical Physics}\ }\textbf {\bibinfo {volume} {159}},\ \bibinfo {pages} {014201} (\bibinfo {year} {2023})}\BibitemShut {NoStop}%
\bibitem [{\citenamefont {Nielsen}\ \emph {et~al.}(2024)\citenamefont {Nielsen}, \citenamefont {Carvalho}, \citenamefont {Goodwin}, \citenamefont {Wili},\ and\ \citenamefont {Nielsen}}]{PLATO_adv}%
  \BibitemOpen
  \bibfield  {author} {\bibinfo {author} {\bibfnamefont {A.~B.}\ \bibnamefont {Nielsen}}, \bibinfo {author} {\bibfnamefont {J.~P.~A.}\ \bibnamefont {Carvalho}}, \bibinfo {author} {\bibfnamefont {D.~L.}\ \bibnamefont {Goodwin}}, \bibinfo {author} {\bibfnamefont {N.}~\bibnamefont {Wili}},\ and\ \bibinfo {author} {\bibfnamefont {N.~C.}\ \bibnamefont {Nielsen}},\ }\bibfield  {title} {\enquote {\bibinfo {title} {Dynamic nuclear polarization pulse sequence engineering using single-spin vector effective hamiltonians},}\ }\href@noop {} {\bibfield  {journal} {\bibinfo  {journal} {Phys. Chem. Chem. Phys.}\ }\textbf {\bibinfo {volume} {26}},\ \bibinfo {pages} {28208--28219} (\bibinfo {year} {2024})}\BibitemShut {NoStop}%
\bibitem [{\citenamefont {Vinding}\ \emph {et~al.}(2013)\citenamefont {Vinding}, \citenamefont {Laustsen}, \citenamefont {Maximov}, \citenamefont {S{\o}gaard}, \citenamefont {Ardenkj{\ae}r-Larsen},\ and\ \citenamefont {Nielsen}}]{OC_MRI_DNP}%
  \BibitemOpen
  \bibfield  {author} {\bibinfo {author} {\bibfnamefont {M.~S.}\ \bibnamefont {Vinding}}, \bibinfo {author} {\bibfnamefont {C.}~\bibnamefont {Laustsen}}, \bibinfo {author} {\bibfnamefont {I.~I.}\ \bibnamefont {Maximov}}, \bibinfo {author} {\bibfnamefont {L.~V.}\ \bibnamefont {S{\o}gaard}}, \bibinfo {author} {\bibfnamefont {J.~H.}\ \bibnamefont {Ardenkj{\ae}r-Larsen}},\ and\ \bibinfo {author} {\bibfnamefont {N.~C.}\ \bibnamefont {Nielsen}},\ }\bibfield  {title} {\enquote {\bibinfo {title} {Dynamic nuclear polarization and optimal control spatial-selective 13c mri and mrs},}\ }\href@noop {} {\bibfield  {journal} {\bibinfo  {journal} {Journal of Magnetic Resonance}\ }\textbf {\bibinfo {volume} {227}},\ \bibinfo {pages} {57--61} (\bibinfo {year} {2013})}\BibitemShut {NoStop}%
\bibitem [{\citenamefont {Carvalho}\ \emph {et~al.}(2025{\natexlab{a}})\citenamefont {Carvalho}, \citenamefont {Goodwin}, \citenamefont {Wili}, \citenamefont {Nielsen},\ and\ \citenamefont {Nielsen}}]{OC_DNP}%
  \BibitemOpen
  \bibfield  {author} {\bibinfo {author} {\bibfnamefont {J.~P.}\ \bibnamefont {Carvalho}}, \bibinfo {author} {\bibfnamefont {D.~L.}\ \bibnamefont {Goodwin}}, \bibinfo {author} {\bibfnamefont {N.}~\bibnamefont {Wili}}, \bibinfo {author} {\bibfnamefont {A.~B.}\ \bibnamefont {Nielsen}},\ and\ \bibinfo {author} {\bibfnamefont {N.~C.}\ \bibnamefont {Nielsen}},\ }\bibfield  {title} {\enquote {\bibinfo {title} {Optimal control design strategies for pulsed dynamic nuclear polarization},}\ }\href@noop {} {\bibfield  {journal} {\bibinfo  {journal} {The Journal of Chemical Physics}\ }\textbf {\bibinfo {volume} {162}},\ \bibinfo {pages} {054111} (\bibinfo {year} {2025}{\natexlab{a}})}\BibitemShut {NoStop}%
\bibitem [{\citenamefont {To{\v s}ner}\ \emph {et~al.}(2006)\citenamefont {To{\v s}ner}, \citenamefont {Glaser}, \citenamefont {Khaneja},\ and\ \citenamefont {Nielsen}}]{OCEffHam}%
  \BibitemOpen
  \bibfield  {author} {\bibinfo {author} {\bibfnamefont {Z.}~\bibnamefont {To{\v s}ner}}, \bibinfo {author} {\bibfnamefont {S.~J.}\ \bibnamefont {Glaser}}, \bibinfo {author} {\bibfnamefont {N.}~\bibnamefont {Khaneja}},\ and\ \bibinfo {author} {\bibfnamefont {N.~C.}\ \bibnamefont {Nielsen}},\ }\bibfield  {title} {\enquote {\bibinfo {title} {Effective hamiltonians by optimal control: Solid-state nmr double-quantum planar and isotropic dipolar recoupling},}\ }\href@noop {} {\bibfield  {journal} {\bibinfo  {journal} {The Journal of Chemical Physics}\ }\textbf {\bibinfo {volume} {125}},\ \bibinfo {pages} {184502} (\bibinfo {year} {2006})}\BibitemShut {NoStop}%
\bibitem [{\citenamefont {Vega}(1978)}]{fictitious_vega}%
  \BibitemOpen
  \bibfield  {author} {\bibinfo {author} {\bibfnamefont {S.}~\bibnamefont {Vega}},\ }\bibfield  {title} {\enquote {\bibinfo {title} {{Fictitious spin 1/2 operator formalism for multiple quantum NMR}},}\ }\href@noop {} {\bibfield  {journal} {\bibinfo  {journal} {The Journal of Chemical Physics}\ }\textbf {\bibinfo {volume} {68}},\ \bibinfo {pages} {5518--5527} (\bibinfo {year} {1978})}\BibitemShut {NoStop}%
\bibitem [{\citenamefont {Wokaun}\ and\ \citenamefont {Ernst}(1977)}]{fictitious_wokaun_ernst}%
  \BibitemOpen
  \bibfield  {author} {\bibinfo {author} {\bibfnamefont {A.}~\bibnamefont {Wokaun}}\ and\ \bibinfo {author} {\bibfnamefont {R.~R.}\ \bibnamefont {Ernst}},\ }\bibfield  {title} {\enquote {\bibinfo {title} {{Selective excitation and detection in multilevel spin systems: Application of single transition operators}},}\ }\href@noop {} {\bibfield  {journal} {\bibinfo  {journal} {The Journal of Chemical Physics}\ }\textbf {\bibinfo {volume} {67}},\ \bibinfo {pages} {1752--1758} (\bibinfo {year} {1977})}\BibitemShut {NoStop}%
\bibitem [{\citenamefont {Hartmann}\ and\ \citenamefont {Hahn}(1962)}]{HartmanHahn}%
  \BibitemOpen
  \bibfield  {author} {\bibinfo {author} {\bibfnamefont {S.}~\bibnamefont {Hartmann}}\ and\ \bibinfo {author} {\bibfnamefont {E.}~\bibnamefont {Hahn}},\ }\bibfield  {title} {\enquote {\bibinfo {title} {Nuclear double resonance in the rotating frame},}\ }\href@noop {} {\bibfield  {journal} {\bibinfo  {journal} {Physical Review}\ }\textbf {\bibinfo {volume} {128}},\ \bibinfo {pages} {2042} (\bibinfo {year} {1962})}\BibitemShut {NoStop}%
\bibitem [{\citenamefont {Carvalho}\ \emph {et~al.}(2025{\natexlab{b}})\citenamefont {Carvalho}, \citenamefont {Nielsen}, \citenamefont {Balig{\'a}cs}, \citenamefont {Wili},\ and\ \citenamefont {Nielsen}}]{DNP_SSNMR_Carvalho:2025aa}%
  \BibitemOpen
  \bibfield  {author} {\bibinfo {author} {\bibfnamefont {J.}~\bibnamefont {Carvalho}}, \bibinfo {author} {\bibfnamefont {A.~B.}\ \bibnamefont {Nielsen}}, \bibinfo {author} {\bibfnamefont {E.}~\bibnamefont {Balig{\'a}cs}}, \bibinfo {author} {\bibfnamefont {N.}~\bibnamefont {Wili}},\ and\ \bibinfo {author} {\bibfnamefont {N.~C.}\ \bibnamefont {Nielsen}},\ }\bibfield  {title} {\enquote {\bibinfo {title} {Bridging dynamic nuclear polarization and solid-state nmr dipolar recoupling: From static single crystal to spinning powders},}\ }\href@noop {} {\bibfield  {journal} {\bibinfo  {journal} {The Journal of Physical Chemistry Letters}\ ,\ \bibinfo {pages} {4363--4371}} (\bibinfo {year} {2025}{\natexlab{b}})}\BibitemShut {NoStop}%
\bibitem [{\citenamefont {Nelder}\ and\ \citenamefont {Mead}(1965)}]{simplex}%
  \BibitemOpen
  \bibfield  {author} {\bibinfo {author} {\bibfnamefont {J.~A.}\ \bibnamefont {Nelder}}\ and\ \bibinfo {author} {\bibfnamefont {R.}~\bibnamefont {Mead}},\ }\bibfield  {title} {\enquote {\bibinfo {title} {{A Simplex Method for Function Minimization}},}\ }\href@noop {} {\bibfield  {journal} {\bibinfo  {journal} {The Computer Journal}\ }\textbf {\bibinfo {volume} {7}},\ \bibinfo {pages} {308--313} (\bibinfo {year} {1965})}\BibitemShut {NoStop}%
\bibitem [{\citenamefont {MATLAB}(2024)}]{MATLAB:2010}%
  \BibitemOpen
  \bibfield  {author} {\bibinfo {author} {\bibnamefont {MATLAB}},\ }\href@noop {} {\emph {\bibinfo {title} {version 24.2.0.2773142 (R2024b)}}}\ (\bibinfo  {publisher} {The MathWorks Inc.},\ \bibinfo {address} {Natick, Massachusetts},\ \bibinfo {year} {2024})\BibitemShut {NoStop}%
\bibitem [{\citenamefont {Levitt}(1986)}]{LEVITT_compositepulses}%
  \BibitemOpen
  \bibfield  {author} {\bibinfo {author} {\bibfnamefont {M.~H.}\ \bibnamefont {Levitt}},\ }\bibfield  {title} {\enquote {\bibinfo {title} {Composite pulses},}\ }\href@noop {} {\bibfield  {journal} {\bibinfo  {journal} {Progress in Nuclear Magnetic Resonance Spectroscopy}\ }\textbf {\bibinfo {volume} {18}},\ \bibinfo {pages} {61--122} (\bibinfo {year} {1986})}\BibitemShut {NoStop}%
\bibitem [{\citenamefont {Levitt}, \citenamefont {Freeman},\ and\ \citenamefont {Frenkiel}(1983)}]{LEVITT198347}%
  \BibitemOpen
  \bibfield  {author} {\bibinfo {author} {\bibfnamefont {M.~H.}\ \bibnamefont {Levitt}}, \bibinfo {author} {\bibfnamefont {R.}~\bibnamefont {Freeman}},\ and\ \bibinfo {author} {\bibfnamefont {T.}~\bibnamefont {Frenkiel}},\ }\bibfield  {title} {\enquote {\bibinfo {title} {Broadband decoupling in high-resolution nuclear magnetic resonance spectroscopy},}\ \ }(\bibinfo  {publisher} {Academic Press},\ \bibinfo {year} {1983})\ pp.\ \bibinfo {pages} {47--110}\BibitemShut {NoStop}%
\bibitem [{\citenamefont {Burum}\ and\ \citenamefont {Rhim}(1979)}]{Rhim3}%
  \BibitemOpen
  \bibfield  {author} {\bibinfo {author} {\bibfnamefont {D.~P.}\ \bibnamefont {Burum}}\ and\ \bibinfo {author} {\bibfnamefont {W.~K.}\ \bibnamefont {Rhim}},\ }\bibfield  {title} {\enquote {\bibinfo {title} {Analysis of multiple pulse nmr in solids. iii},}\ }\href@noop {} {\bibfield  {journal} {\bibinfo  {journal} {The Journal of Chemical Physics}\ }\textbf {\bibinfo {volume} {71}},\ \bibinfo {pages} {944--956} (\bibinfo {year} {1979})}\BibitemShut {NoStop}%
\bibitem [{\citenamefont {Burum}, \citenamefont {Linder},\ and\ \citenamefont {Ernst}(1981)}]{BURUM1981173}%
  \BibitemOpen
  \bibfield  {author} {\bibinfo {author} {\bibfnamefont {D.}~\bibnamefont {Burum}}, \bibinfo {author} {\bibfnamefont {M.}~\bibnamefont {Linder}},\ and\ \bibinfo {author} {\bibfnamefont {R.}~\bibnamefont {Ernst}},\ }\bibfield  {title} {\enquote {\bibinfo {title} {Low-power multipulse line narrowing in solid-state nmr},}\ }\href@noop {} {\bibfield  {journal} {\bibinfo  {journal} {Journal of Magnetic Resonance (1969)}\ }\textbf {\bibinfo {volume} {44}},\ \bibinfo {pages} {173--188} (\bibinfo {year} {1981})}\BibitemShut {NoStop}%
\bibitem [{\citenamefont {Hohwy}\ and\ \citenamefont {Nielsen}(1997)}]{MSHOT3}%
  \BibitemOpen
  \bibfield  {author} {\bibinfo {author} {\bibfnamefont {M.}~\bibnamefont {Hohwy}}\ and\ \bibinfo {author} {\bibfnamefont {N.~C.}\ \bibnamefont {Nielsen}},\ }\bibfield  {title} {\enquote {\bibinfo {title} {Elimination of high order terms in multiple pulse nuclear magnetic resonance spectroscopy: Application to homonuclear decoupling in solids},}\ }\href@noop {} {\bibfield  {journal} {\bibinfo  {journal} {The Journal of Chemical Physics}\ }\textbf {\bibinfo {volume} {106}},\ \bibinfo {pages} {7571--7586} (\bibinfo {year} {1997})}\BibitemShut {NoStop}%
\bibitem [{\citenamefont {Equbal}\ \emph {et~al.}(2015)\citenamefont {Equbal}, \citenamefont {Bjerring}, \citenamefont {Madhu},\ and\ \citenamefont {Nielsen}}]{UHDS}%
  \BibitemOpen
  \bibfield  {author} {\bibinfo {author} {\bibfnamefont {A.}~\bibnamefont {Equbal}}, \bibinfo {author} {\bibfnamefont {M.}~\bibnamefont {Bjerring}}, \bibinfo {author} {\bibfnamefont {P.~K.}\ \bibnamefont {Madhu}},\ and\ \bibinfo {author} {\bibfnamefont {N.~C.}\ \bibnamefont {Nielsen}},\ }\bibfield  {title} {\enquote {\bibinfo {title} {A unified heteronuclear decoupling strategy for magic-angle-spinning solid-state nmr spectroscopy},}\ }\href@noop {} {\bibfield  {journal} {\bibinfo  {journal} {The Journal of Chemical Physics}\ }\textbf {\bibinfo {volume} {142}},\ \bibinfo {pages} {184201} (\bibinfo {year} {2015})}\BibitemShut {NoStop}%
\bibitem [{\citenamefont {Mote}, \citenamefont {Agarwal},\ and\ \citenamefont {Madhu}(2016)}]{MOTE20161}%
  \BibitemOpen
  \bibfield  {author} {\bibinfo {author} {\bibfnamefont {K.~R.}\ \bibnamefont {Mote}}, \bibinfo {author} {\bibfnamefont {V.}~\bibnamefont {Agarwal}},\ and\ \bibinfo {author} {\bibfnamefont {P.}~\bibnamefont {Madhu}},\ }\bibfield  {title} {\enquote {\bibinfo {title} {Five decades of homonuclear dipolar decoupling in solid-state nmr: Status and outlook},}\ }\href@noop {} {\bibfield  {journal} {\bibinfo  {journal} {Progress in Nuclear Magnetic Resonance Spectroscopy}\ }\textbf {\bibinfo {volume} {97}},\ \bibinfo {pages} {1--39} (\bibinfo {year} {2016})}\BibitemShut {NoStop}%
\bibitem [{\citenamefont {Ed{\'e}n}\ and\ \citenamefont {Levitt}(1999)}]{Eden_decoupling}%
  \BibitemOpen
  \bibfield  {author} {\bibinfo {author} {\bibfnamefont {M.}~\bibnamefont {Ed{\'e}n}}\ and\ \bibinfo {author} {\bibfnamefont {M.~H.}\ \bibnamefont {Levitt}},\ }\bibfield  {title} {\enquote {\bibinfo {title} {Pulse sequence symmetries in the nuclear magnetic resonance of spinning solids: Application to heteronuclear decoupling},}\ }\href@noop {} {\bibfield  {journal} {\bibinfo  {journal} {The Journal of Chemical Physics}\ }\textbf {\bibinfo {volume} {111}},\ \bibinfo {pages} {1511--1519} (\bibinfo {year} {1999})}\BibitemShut {NoStop}%
\bibitem [{\citenamefont {Lee}\ \emph {et~al.}(1995)\citenamefont {Lee}, \citenamefont {Kurur}, \citenamefont {Helmle}, \citenamefont {Johannessen}, \citenamefont {Nielsen},\ and\ \citenamefont {Levitt}}]{C7}%
  \BibitemOpen
  \bibfield  {author} {\bibinfo {author} {\bibfnamefont {Y.}~\bibnamefont {Lee}}, \bibinfo {author} {\bibfnamefont {N.}~\bibnamefont {Kurur}}, \bibinfo {author} {\bibfnamefont {M.}~\bibnamefont {Helmle}}, \bibinfo {author} {\bibfnamefont {O.}~\bibnamefont {Johannessen}}, \bibinfo {author} {\bibfnamefont {N.}~\bibnamefont {Nielsen}},\ and\ \bibinfo {author} {\bibfnamefont {M.}~\bibnamefont {Levitt}},\ }\bibfield  {title} {\enquote {\bibinfo {title} {Efficient dipolar recoupling in the nmr of rotating solids. a sevenfold symmetric radiofrequency pulse sequence},}\ }\href@noop {} {\bibfield  {journal} {\bibinfo  {journal} {Chemical Physics Letters}\ }\textbf {\bibinfo {volume} {242}},\ \bibinfo {pages} {304--309} (\bibinfo {year} {1995})}\BibitemShut {NoStop}%
\bibitem [{\citenamefont {Hohwy}\ \emph {et~al.}(1998)\citenamefont {Hohwy}, \citenamefont {Jakobsen}, \citenamefont {Ed{\'e}n}, \citenamefont {Levitt},\ and\ \citenamefont {Nielsen}}]{POSTC7}%
  \BibitemOpen
  \bibfield  {author} {\bibinfo {author} {\bibfnamefont {M.}~\bibnamefont {Hohwy}}, \bibinfo {author} {\bibfnamefont {H.~J.}\ \bibnamefont {Jakobsen}}, \bibinfo {author} {\bibfnamefont {M.}~\bibnamefont {Ed{\'e}n}}, \bibinfo {author} {\bibfnamefont {M.~H.}\ \bibnamefont {Levitt}},\ and\ \bibinfo {author} {\bibfnamefont {N.~C.}\ \bibnamefont {Nielsen}},\ }\bibfield  {title} {\enquote {\bibinfo {title} {Broadband dipolar recoupling in the nuclear magnetic resonance of rotating solids: A compensated c7 pulse sequence},}\ }\href@noop {} {\bibfield  {journal} {\bibinfo  {journal} {The Journal of Chemical Physics}\ }\textbf {\bibinfo {volume} {108}},\ \bibinfo {pages} {2686--2694} (\bibinfo {year} {1998})}\BibitemShut {NoStop}%
\bibitem [{\citenamefont {Levitt}(2002)}]{Levitt_symm_rec}%
  \BibitemOpen
  \bibfield  {author} {\bibinfo {author} {\bibfnamefont {M.~H.}\ \bibnamefont {Levitt}},\ }\bibfield  {title} {\enquote {\bibinfo {title} {Symmetry-based pulse sequences in magic-angle spinning solid-state nmr},}\ }in\ \href@noop {} {\emph {\bibinfo {booktitle} {Encyclopedia of Nuclear Magnetic Resonance. Volume 9, Advances in NMR}}},\ \bibinfo {editor} {edited by\ \bibinfo {editor} {\bibfnamefont {D.~M.}\ \bibnamefont {Grant}}\ and\ \bibinfo {editor} {\bibfnamefont {R.~K.}\ \bibnamefont {Harris}}}\ (\bibinfo  {publisher} {Wiley},\ \bibinfo {year} {2002})\ pp.\ \bibinfo {pages} {165--196}\BibitemShut {NoStop}%
\bibitem [{\citenamefont {Doll}\ and\ \citenamefont {Jeschke}(2017)}]{Doll:2017wb}%
  \BibitemOpen
  \bibfield  {author} {\bibinfo {author} {\bibfnamefont {A.}~\bibnamefont {Doll}}\ and\ \bibinfo {author} {\bibfnamefont {G.}~\bibnamefont {Jeschke}},\ }\bibfield  {title} {\enquote {\bibinfo {title} {Wideband frequency-swept excitation in pulsed epr spectroscopy.}}\ }\href {https://doi.org/10.1016/j.jmr.2017.01.004} {\bibfield  {journal} {\bibinfo  {journal} {J Magn Reson}\ }\textbf {\bibinfo {volume} {280}},\ \bibinfo {pages} {46--62} (\bibinfo {year} {2017})}\BibitemShut {NoStop}%
\bibitem [{\citenamefont {Gupta}\ \emph {et~al.}(2015)\citenamefont {Gupta}, \citenamefont {Hou}, \citenamefont {Polenova},\ and\ \citenamefont {Vega}}]{GUPTA201517}%
  \BibitemOpen
  \bibfield  {author} {\bibinfo {author} {\bibfnamefont {R.}~\bibnamefont {Gupta}}, \bibinfo {author} {\bibfnamefont {G.}~\bibnamefont {Hou}}, \bibinfo {author} {\bibfnamefont {T.}~\bibnamefont {Polenova}},\ and\ \bibinfo {author} {\bibfnamefont {A.~J.}\ \bibnamefont {Vega}},\ }\bibfield  {title} {\enquote {\bibinfo {title} {Rf inhomogeneity and how it controls cpmas},}\ }\href@noop {} {\bibfield  {journal} {\bibinfo  {journal} {Solid State Nuclear Magnetic Resonance}\ }\textbf {\bibinfo {volume} {72}},\ \bibinfo {pages} {17--26} (\bibinfo {year} {2015})}\BibitemShut {NoStop}%
\end{thebibliography}%

\end{document}